%% file: main.tex
\documentclass[sigconf,9pt]{acmart}
\usepackage{array}           
\usepackage{colortbl}        
\usepackage{xcolor}          
\usepackage{amssymb}         
\usepackage{amsmath}
\usepackage{tabularx} 
\usepackage{makecell}
\usepackage{float}  
\usepackage{placeins}  

\usepackage{subcaption}
\AtBeginDocument{%
  }

\newcommand{\change}[1]{\textcolor{black}{#1}}

\setcopyright{none}

\newcommand{\sysname}{{GreenScatter}}

\begin{document}

\title{\sysname{}: Through-Canopy Soil Moisture Sensing with UAV-Mounted Radar}

\author{Luke Jacobs}
\affiliation{%
  \institution{University of Illinois Urbana-Champaign}
  \state{Illinois}
  \country{USA}
}
\email{lukedj2@illinois.edu}
\orcid{0000-0002-3722-5508}

\author{Ishfaq Aziz}
\affiliation{%
  \institution{University of Illinois Urbana-Champaign}
  \state{Illinois}
  \country{USA}
}
\email{ishfaqa2@illinois.edu}
\orcid{0000-0001-5723-0331}

\author{Benhao Lu}
\affiliation{%
  \institution{University of Illinois Urbana-Champaign}
  \state{Illinois}
  \country{USA}
}
\email{benhaol2@illinois.edu}
\orcid{0009-0009-4459-9813}

\author{Alireza Tabatabaeenejad}
\affiliation{%
  \institution{The Aerospace Corporation}
  \city{El Segundo}
  \state{California}
  \country{USA}
}
\email{alirezat@aero.org}
\orcid{0000-0002-5098-8526}

\author{Mohamad Alipour}
\affiliation{%
  \institution{University of Illinois Urbana-Champaign}
  \state{Illinois}
  \country{USA}
}
\email{alipour@illinois.edu}
\orcid{0000-0003-2018-134X}

\author{Elahe Soltanaghai}
\affiliation{%
  \institution{University of Illinois Urbana-Champaign}
  \state{Illinois}
  \country{USA}
}
\email{elahe@illinois.edu}
\orcid{0009-0006-5040-5438}








\renewcommand{\shortauthors}{Jacobs et al.}

\begin{abstract}
  Soil moisture is a critical variable for managing irrigation, improving crop yield, and understanding field-scale hydrology. Radars mounted on unmanned aerial vehicles (UAVs) offer a promising means to monitor soil moisture over large fields with flexible, high-resolution coverage. However, during the growing season, canopy scattering and soil reflections become strongly coupled in the radar measurement. These coupled effects vary with crop structure or flight altitude, complicating the retrieval of soil moisture. To overcome this challenge, we present \sysname{}, a physics-based soil moisture retrieval framework for nadir-looking wideband UAV radars. \sysname{} introduces a microwave radiative transfer model that explicitly captures the dominant electromagnetic interactions between vegetation and soil, enabling accurate modeling of coherent ground backscatter through canopy. In parallel, it develops a radar cross-section (RCS) estimation method that transforms time-domain radar signals into calibrated wideband RCS spectra, isolating soil reflections while compensating for hardware and waveform effects. Together, these components enable robust soil moisture estimation through vegetation across varying canopy conditions and UAV configurations. Field experiments across multiple corn and soybean sites demonstrate consistent retrieval with an average volumetric water content (VWC) error of 4.49\%.
\end{abstract}





\maketitle

\input{sections/1-introduction}
\input{sections/2-relatedwork}

\input{sections/3-methodology}

\input{sections/4-implementation}

\input{sections/5-evaluation}
\input{sections/6-discussion}
\input{sections/7-conclusion}

\begin{acks}
The paper was partially funded by NASA Award \#80NSSC25K7635 and NSF Awards \#2434387 as well as NSF Graduate Research Fellowship Program (GRFP). We also thank Keysight Technology for hardware support, and UIUC Prairie Research Institute – Center for Digital Agriculture (PRI–CDA) Joint Research Program for supporting our field experimentation. We are grateful to Mingyue Tang, Yuxiang  Zhao, Zixin Wang, and Tianyi Zhong who assisted in conducting the field experiments. 
\end{acks}


\bibliographystyle{ACM-Reference-Format}
\bibliography{sample-base}

\end{document}

%% file: sections/1-introduction.tex
\section{Introduction}

\begin{figure}[t]
\centering
\includegraphics[width=1.0\linewidth]{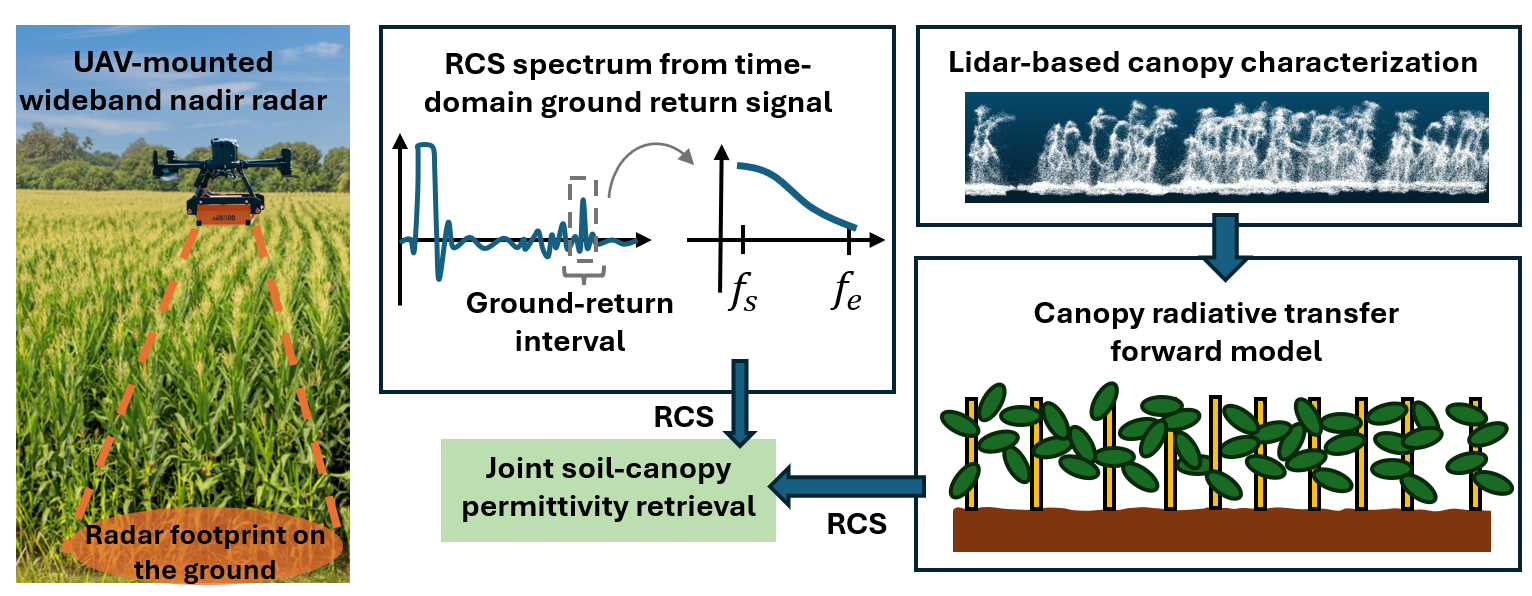}
\caption{\sysname{} offers robust soil moisture sensing through canopy using UAV-mounted radar}
\label{fig:concept}
\end{figure}

Many agricultural decisions throughout the growing season rely on soil moisture monitoring, from scheduling irrigation to determining the best times for fertilizer and pesticide application. Yet, obtaining soil moisture data across large fields remains challenging. Manual in situ sampling is labor-intensive, and dense sensor networks are costly, limiting widespread adoption of data-driven farming. Drone-based remote sensing technologies using optical or radio frequency (RF) sensors offer farmers high-resolution mapping of fields and crop canopies, 
but estimating soil moisture beneath vegetation remains difficult, especially as crop structure evolves throughout the growing season. The challenge is that optical sensors capture only the uppermost surface information and cannot sense soil moisture through the canopy. RF systems such as synthetic aperture radar~(SAR) and ground-penetrating radar (GPR) can penetrate vegetation, but canopy-induced scattering and attenuation distort the soil reflection, complicating the separation of vegetation and soil contributions. Moreover, UAV altitude fluctuations and platform vibrations introduce additional variability in signal attenuation, further degrading the reliability of soil reflection measurements.


\change{Previous microwave remote sensing studies have attempted to address these challenges using calibrated polarimetric SAR systems in side-looking configurations~\cite{lamichhane2025soil}. In such systems, vegetation volume scattering mainly appears in the cross-polarized channels, while soil backscatter is primarily observed in the co-polarized channels, allowing partial separation of canopy and soil contributions. However, at side-looking incidence angles, specular reflections from the soil surface are directed away from the radar, causing canopy scattering to dominate the received signal and reducing the sensitivity to soil moisture.
More recent research has explored the use of ground-penetrating radar in nadir-looking configurations, which provide stronger soil reflections at the cost of losing the polarimetric data. 
These methods typically estimate soil moisture using surface reflection characteristics such as amplitude \cite{pramudita_soil_2022}, time-of-flight \cite{aziz_bayesian_2024, josephson_low-cost_2021}, or relative phase across antennas \cite{khan_estimating_2022}. However, all of these parameters are simultaneously influenced by drone altitude, platform motion, canopy scattering, and soil properties, which complicates their direct use in UAV-based sensing. Consequently, most existing approaches are either tested in fixed experimental setups \cite{khan_estimating_2022} or rely on manual canopy characterization \cite{pramudita_soil_2022}, limiting their practicality for large-scale field deployment.}

To overcome these limitations, we introduce \sysname{}, a new soil moisture retrieval framework that fuses data from a nadir-looking wideband radar and LiDAR to jointly characterize soil and vegetation properties. \sysname{}'s key insight lies in the radar’s geometry: when viewed from nadir, the radar captures strong specular reflections from the ground instead of losing them off to the side. This dramatically boosts the signal-to-clutter ratio and allows soil reflections to emerge more clearly even through the canopy. Building on this geometry, we develop a radiative transfer–based retrieval algorithm that models how wideband, nadir-looking radar signals are scattered, absorbed, and transmitted through vegetation before reflecting from the soil surface. The model jointly accounts for canopy–soil interactions and variations in drone altitude, ensuring a physically consistent representation of the observed radar signal. \change{To support the radiative transfer model, we compute a hardware-independent ground radar cross section (RCS) that reflects soil permittivity. Higher moisture content produces stronger reflections due to increased dielectric contrast at the air–soil interface. LiDAR data is then used to derive canopy structural parameters such as height and density, providing the contextual information needed to model and invert canopy attenuation effects.} As such, \sysname{} consists of three main modules:

\vspace{.3em}\noindent \textbf{Radiative Transfer Model for Downward-Facing Radar:} We adapt a microwave radiative transfer model proposed by \cite{burgin_generalized_2011} to nadir-looking, wideband radars. Unlike side-looking radar configurations where vegetation volume scattering and surface–trunk interactions dominate, the nadir-looking configuration introduces a strong coherent component associated with the specular ground reflection. We extend an existing soil scattering model \cite{tabatabaeenejad_bistatic_2006} to explicitly incorporate this coherent term, enabling accurate modeling of the surface backscatter at nadir incidence. Furthermore, the low altitude and wide beamwidth of UAV-mounted radar systems illuminate the ground over a broad range of incidence angles, each exhibiting varying soil backscatter strengths and canopy attenuation. \sysname{} addresses this complexity by computing an effective radar footprint that integrates both coherent and incoherent backscatter contributions. \change{The resulting model establishes the relationship between soil moisture, canopy permittivity, and the resulting RCS. We then use this model together with measured RCS values to invert for soil moisture.}



\vspace{.3em}\noindent \textbf{Radar Frequency Response \& Ground RCS Estimation:} 
A major challenge in linking raw radar data to the radiative transfer model lies in the mismatch between what radars measure and what the model represents. Commercial wideband radars record time-domain reflections that include the transmitted pulse waveform and overlapping returns from multiple scatterers such as vegetation, ground, and nearby objects. In contrast, radiative transfer models operate in the frequency domain, using Maxwell’s equations to describe how vegetation, soil, and other targets contribute to the ground radar cross section (RCS) at each frequency. \change{Bridging these representations is challenging, because of radar’s frequency-varying transmit power and mixed time-domain responses. To overcome this, 
\sysname{} introduces a signal processing pipeline that first isolates the ground reflection in the time domain using the radar’s high range resolution, then suppresses clutter, and transforms the gated segment into the frequency domain.} A reference target with known RCS is then used to remove antenna pattern effects, waveform shaping, and other hardware-related biases. The resulting calibrated RCS spectrum represents the coherent ground backscatter attenuated by the canopy and serves as the observation for inversion.

\vspace{.3em}\noindent \textbf{LiDAR-Based Canopy Structure Extraction:} \sysname{} utilizes LiDAR data to extract key canopy structural parameters required by the radiative transfer model. Although LiDAR can, in principle, capture detailed canopy geometry, agricultural canopies often produce sparse and incomplete point clouds due to occlusion and limited sampling density, making it difficult to recover fine-scale structural details needed for accurate canopy attenuation modeling. \sysname{} overcomes this limitation by exploiting the inherent row-based organization of crop fields and incorporating prior knowledge of crop morphology (e.g., thick central stems in corn versus denser, shorter structures in soybeans) to robustly estimate plant count and height. For leaf-related parameters, \sysname{} introduces a voxel-based estimation of the leaf area index (LAI) and integrates it with crop-specific allometric relationships to derive leaf density, enabling a physically-grounded yet computationally-efficient canopy representation.

We evaluate the proposed approach using a commercial UAV-mounted ground-penetrating radar (GPR) system operating in the 200–900 MHz frequency range \cite{radar_systems_zond_nodate} across six different agricultural fields cultivated with corn and soybean at varying densities and growth stages. Our results show that the method achieves an average error of 4.49\% in estimating soil volumetric water content, which is sufficient for practical farming decision-making. We further validate the robustness of the system through a 9-day longitudinal experiment that included two rainfall events causing substantial variations in soil permittivity, as well as tests conducted at multiple drone flight altitudes.
As such, our contributions are as follows:
\begin{itemize}
\item We develop a physics-driven radiative transfer model 
for nadir-looking wideband radar configuration. This model serves as the forward model of an inversion-based retrieval algorithm that decouples canopy, soil, and drone location contributions in radar measurements, enabling accurate UAV-based soil moisture estimation through vegetation.
\item We develop a signal transformation framework that converts time-domain radar measurements into a wideband frequency-domain radar cross-section (RCS) spectrum, bridging the gap between radar observations and radiative transfer modeling while accounting for the effective radar footprint. 
\item We introduce a canopy structure parameterization algorithm that estimates key crop structural attributes such as height and density from sparse LiDAR point clouds, providing essential inputs for modeling canopy-induced attenuation.
\item We conducted extensive in-situ experimentation using a commercial UAV-mounted radar across 6 agricultural fields with varying canopy coverage of soy and corn crops. The results indicate an average error of 4.49\% in estimating soil volume water content estimation through canopy with robustness to drone mobility, height, and multipath factors. 
\end{itemize}

%% file: sections/2-relatedwork.tex
\section{Background and Related Work}

\begin{table*}[h]
\centering
\renewcommand{\arraystretch}{1.3} 
\begin{tabular}{l|l|l|l|l|l}
\hline
\textbf{} & \textbf{Sensing Modality} & \textbf{Reader Distance} & \textbf{Probe-Free} & \makecell{\textbf{Canopy Invariance}} & \textbf{UAV-Tested} \\
\hline
\hline GPSoil \cite{dong_gpsoil} & GPS Receiver Tag & N/A & \textcolor{red}{$\times$} & \textcolor{green!60!black}{$\checkmark$} & N/A \\
\hline 
CoMEt \cite{khan_estimating_2022} & SFCW Radar & $< 1$ m & \textcolor{green!60!black}{$\checkmark$} & \textcolor{red}{$\times$} & \textcolor{red}{$\times$} \\
\hline 
Pramudita et al. \cite{pramudita_soil_2022} & SFCW Radar & $< 1$ m & \textcolor{green!60!black}{$\checkmark$} & \textcolor{red}{$\times$} (manual calibration) & \textcolor{green!60!black}{$\checkmark$} \\
\hline
SoilId \cite{ding_soil_2023} & IR-UWB Radar & $<2$ m & \textcolor{red}{$\times$} (buried metal) & \textcolor{red}{$\times$} & \textcolor{green!60!black}{$\checkmark$} \\
\hline 
Josephson et al. \cite{josephson_low-cost_2021} & UWB Radar + Tag & $< 10$ m & \textcolor{green!60!black}{$\checkmark$} & \textcolor{red}{$\times$} & \textcolor{red}{$\times$} \\
\hline 
Wu et al. \cite{wu_new_2019} & Narrowband GPR & $< 5$ m & \textcolor{green!60!black}{$\checkmark$} & \textcolor{red}{$\times$} & \textcolor{green!60!black}{$\checkmark$} \\
\hline 
Vahidi et al. \cite{vahidi_multi-modal_2025} & Wideband GPR & $< 25$ m & \textcolor{green!60!black}{$\checkmark$} & \textcolor{red}{$\times$} (limited to training set) & \textcolor{green!60!black}{$\checkmark$} \\
\hline SoilCares \cite{wang_soilcares} & LoRa TX + RX & $< 80$ m & \textcolor{red}{$\times$} & \textcolor{red}{$\times$} & \textcolor{red}{$\times$} \\
\hline 
\textbf{GreenScatter} & \textbf{Wideband GPR} & \textbf{$\textbf{< 10}$ m} & \textbf{\textcolor{green!60!black}{$\checkmark$}} & \textbf{\textcolor{green!60!black}{$\checkmark$}} & \textbf{\textcolor{green!60!black}{$\checkmark$}} \\
\hline
\end{tabular}
\vspace{.6em}
\caption{\change{Prior work comparison across design dimensions.}}
\label{tab:prior_work_comparison}
\end{table*}

Prior work in soil moisture estimation can be grouped into two categories, polarimetric approaches and GPR-centric approaches, neither of which provide a practical solution to canopy-soil coupling. \change{A comparison of \sysname{} to prior work is provided in Table~\ref{tab:prior_work_comparison}.}

\subsection{Polarimetric-centric Soil Sensing} 

SAR mapping of surface soil moisture under natural vegetation covers has been a key challenge in remote sensing due to strong vegetation effects masking the soil moisture signal by scattering from branches and attenuation by leaves and branches \cite{karam1992microwave}. Prior works have used parametric models for forward scattering \cite{moghaddam2000estimating}, inversion of a physical forward model \cite{jagdhuber2012soil}, regression models \cite{bourgeau2013evaluation} or polarimetric decomposition \cite{hajnsek2003inversion} on L-, P-, and C-band airborne SAR data. However, most of these algorithms are tested in non-forest and non-agricultural lands with thinly vegetated surfaces.

Polarimetric soil moisture retrieval is the dominant line of work in satellite remote sensing \cite{fluhrer_soil_2024, qian_multi-layer_2025, kurum_surface_2021, truong-loi_soil_2015, tabatabaeenejad_potential_2012, zhao_multifrequency_2025}. These methods use side-looking, narrow-band backscatter measurements (typically in L– or P–band) and rely on polarization diversity to separate canopy and soil contributions. Prior studies either apply matrix decomposition \cite{fluhrer_soil_2024} or explicitly parameterize volume and dihedral scattering terms (e.g., trunk–ground and branch–ground interactions) \cite{kurum_surface_2021, truong-loi_soil_2015, tabatabaeenejad_potential_2012}. In practice, however, canopy and soil scattering remain tightly coupled even in polarimetric observations, so both lines of work suffer from a large parameter space being inferred from very limited observables, resulting in an ill-posed inversion where multiple parameter sets can explain the same measurement. More recent work \cite{zhao_multifrequency_2025} incorporates hyperspectral data from a second satellite to constrain canopy structure, but multi-satellite fusion increases revisit time and is therefore not suitable for farming applications. 

\subsection{GPR-centric Soil Sensing}
GPR based approaches \cite{aziz_bayesian_2024, ding_soil_2023, josephson_low-cost_2021, pramudita_soil_2022, lambot_modeling_2004, vahidi_multi-modal_2025, wu_new_2019, khan_estimating_2022} use time domain radar measurements to exploit range resolution. This makes it easier to separate canopy returns from the ground reflection, at the cost of losing polarimetric data. The two most common signal features extracted from these time-domain measurements are time-difference-of-arrival (TDoA) values and reflection amplitudes. 

TDoA based methods \cite{aziz_bayesian_2024, ding_soil_2023, josephson_low-cost_2021, khan_estimating_2022} estimate soil moisture from the delay between surface and subsurface reflections. A central challenge in this approach is isolating the contributions of individual soil layers. One line of work attempts to address this by placing reflectors at known depths \cite{aziz_bayesian_2024, ding_soil_2023, josephson_low-cost_2021}. However, covering a large field would require many reflectors which is infeasible for routine farming scale use. Another line of work avoids buried reflectors by using large antenna arrays and very wide bandwidth to resolve subsurface time delays from natural in situ reflections \cite{khan_estimating_2022}. However, it requires physically large arrays placed near the surface which is incompatible with small UAV platforms or through-canopy sensing.


Reflection amplitude methods provide a third strategy \cite{cheng_estimation_2023, pramudita_soil_2022, lambot_modeling_2004}. In these methods the magnitude of the ground peak in the time trace is used as a proxy for the reflection coefficient which depends on soil moisture. The difficulty is that this same amplitude is also attenuated by the canopy before reaching the soil and again on the way back, so the peak does not uniquely reflect soil conditions. Prior work has tried to compensate for this using manual canopy attenuation calibration \cite{pramudita_soil_2022}, but it does not work during the growing season.


\subsection{LiDAR-Based Canopy Characterization} 
LiDAR has been used to estimate crop structure in a wide range of studies. Earlier work focused on fairly controlled environments where static laser scanners or structure-from-motion systems captured clean point clouds, making it possible to measure basic plant attributes such as height, width and approximate plant volume \cite{Paulus2019Measuring}. More recent efforts explore UAV and mobile LiDAR for full-field mapping of canopy shape (for example plant height, canopy area and spacing) \cite{LiDARReview2024, Choi2023LiDARCrop}. However, these results are mostly shown on early growth stages, where plants are small, inter-row occlusion is limited, and points are easier to interpret. A smaller set of studies also estimate leaf-related parameters such as leaf area index using voxel-based or gap-fraction models originally developed for forestry \cite{Haluboka2021, Wang2020Review, Beland2021, Nguyen2022, Wang2023Profile}. But none of these methods have demonstrated reliability for dense, mature crops where canopy closure causes heavy occlusion and depth ambiguity. Our goal is to develop a canopy parameter extraction pipeline that remains robust under these real-field, mature-canopy conditions.


%% file: sections/3-methodology.tex
\section{\sysname{} System Overview}
\sysname{} retrieves soil moisture by jointly estimating canopy and soil permittivity from UAV radar and LiDAR data. The central idea is to leverage nadir looking wideband radar where the specular ground reflection remains strong and temporally separable from canopy scattering. This makes the ground radar cross section a cleaner observable than in side looking polarimetric systems and reduces the number of unknowns for which the inversion must solve. \sysname{} builds on this by introducing a forward model tailored to nadir wideband radars that links soil radar cross section to soil permittivity while incorporating canopy attenuation driven by canopy structure.

\sysname{}'s workflow is as follows. For each radar acquisition, \sysname{} isolates the ground return in the time domain, converts it into the frequency domain and standardizes it into an RCS spectrum compatible with the forward model. In parallel, LiDAR scans extract canopy structural descriptors such as canopy height and vegetation density, which parameterize the attenuation term in the model. An iterative inversion then matches the measured RCS spectrum with the simulated spectrum across frequency, yielding a joint estimate of canopy and soil permittivity, and thereby soil moisture. The following sections expand on the details of the \sysname{} radiative transfer model, the radar signal processing pipeline, and the LiDAR based canopy characterization.

\subsection{Radiative Transfer Model for Downward-Facing Radar}

\begin{figure}
    \centering
    
    \begin{subfigure}{0.49\linewidth}
        \centering
        \includegraphics[width=\linewidth]{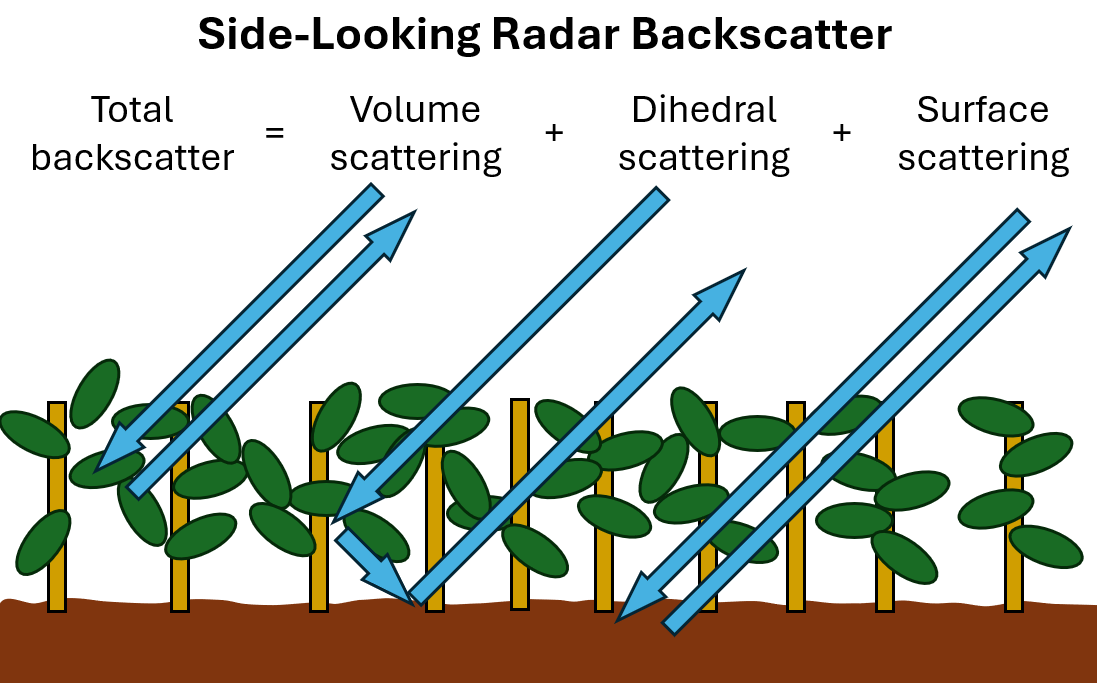}
        \caption{}
        \label{fig:side_looking_backscatter}
    \end{subfigure}
    \hfill
    \begin{subfigure}{0.49\linewidth}
        \centering
        \includegraphics[width=\linewidth]{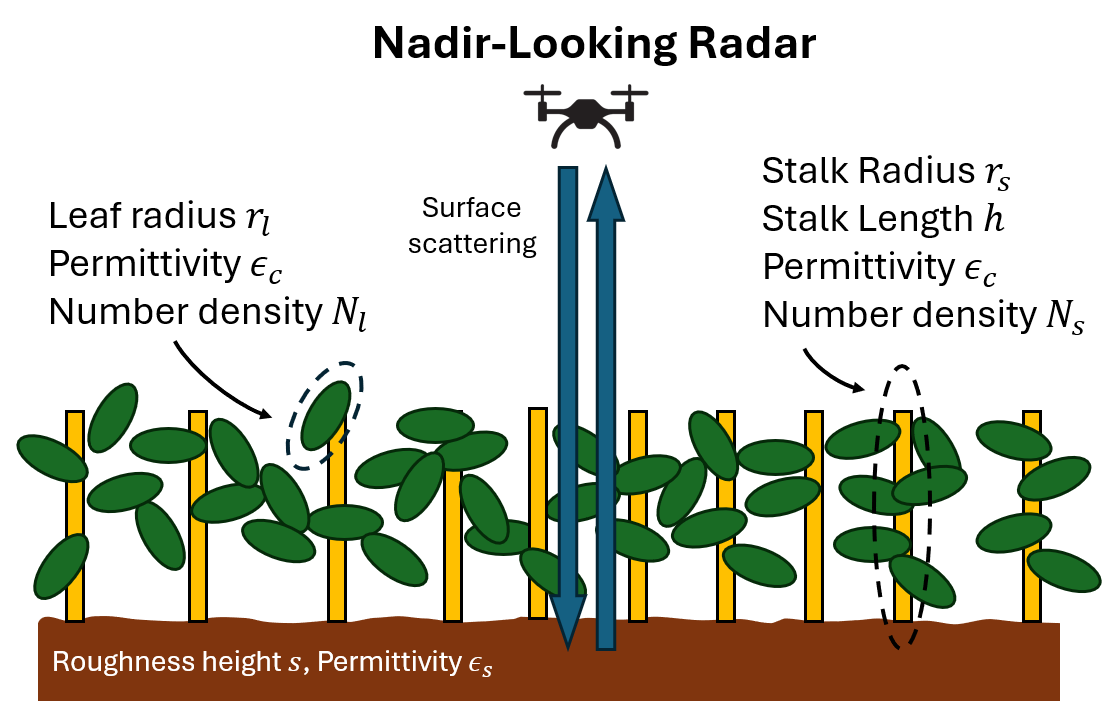}
        \caption{}
        \label{fig:nadir_looking_backscatter}
    \end{subfigure}
    
    \caption{Side-looking (a) backscatter components vs. nadir-looking (b) backscatter components.}
    \label{fig:backscatter_sidelooking_vs_nadirlooking}
\end{figure}

\sysname{} builds on a physically-grounded forward model of RF interactions in agricultural fields, and performs inversion against this model to retrieve soil moisture. Prior physical modeling approaches often rely on radiative transfer theory \cite{tabatabaeenejad2011potential,lang1983electromagnetic, liao2016multiple} to model RF interactions at a side-looking angle. The fundamental challenge with inverting side-looking measurements, however, is that the strong volume and dihedral scattering contributions must be removed from them, which imposes too many unknown variables compared to observations (Figure~\ref{fig:backscatter_sidelooking_vs_nadirlooking}(a)). The key insight in \sysname{} is that for nadir-looking radars, the coherent ground reflection dominates the radar return. Therefore, rather than modeling canopy volume and dihedral scattering, it is only necessary to model the attenuation due to penetration through canopy for the specular ground reflection. We therefore adopt only the attenuation model of \cite{burgin_generalized_2011}, which reduces the number of required parameters while preserving the physics that matters for robust inversion. 

\begin{figure*}[t]
    \centering
    \begin{minipage}{0.33\textwidth}
    \includegraphics[width=.9\linewidth]{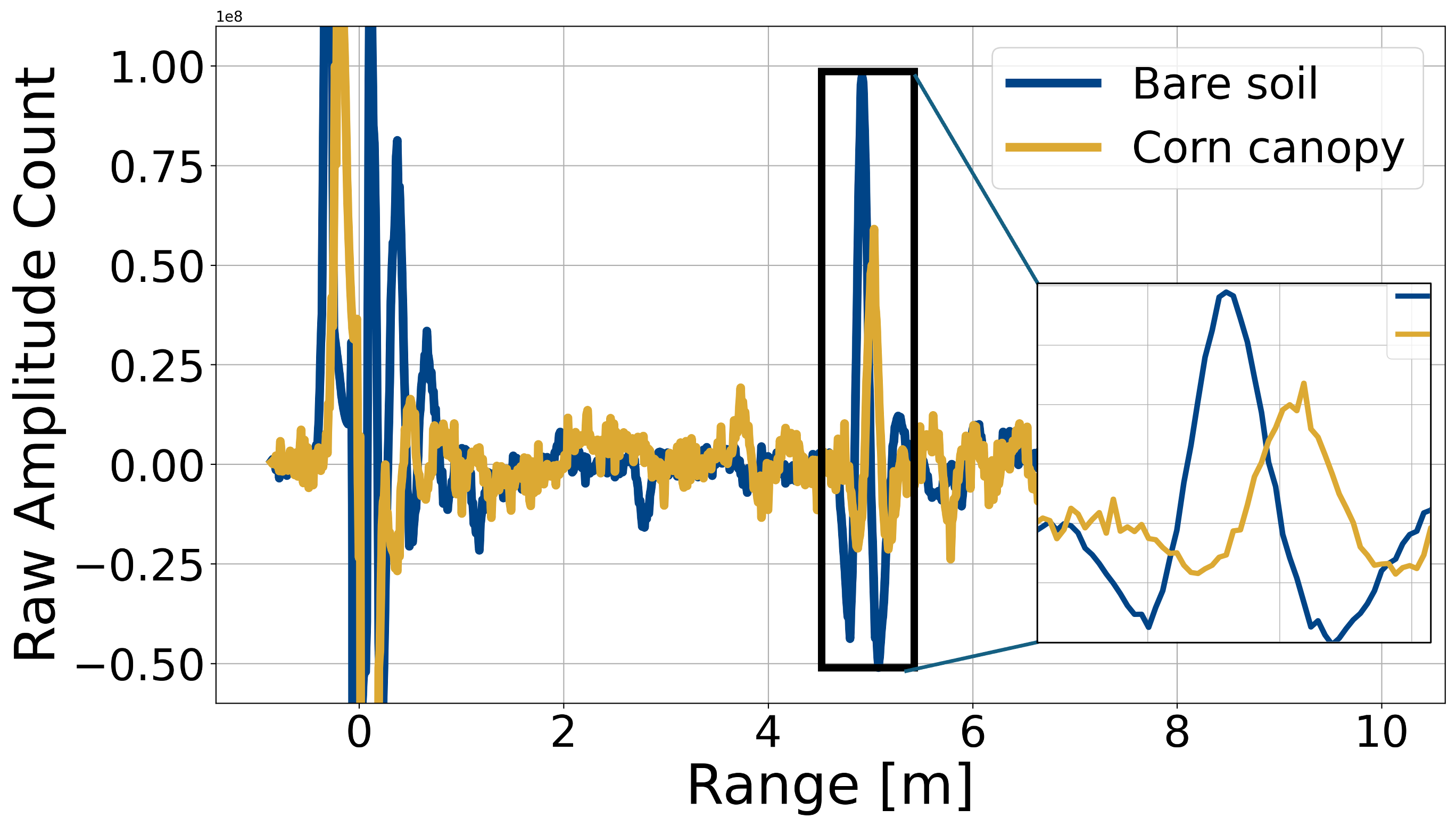}
              \captionsetup{width=.9\linewidth}

    \caption{Comparison between two scans collected at same drone altitudes and moisture but varying canopy coverage}
    \label{fig:canopy_attenuation}
    \end{minipage}
    \begin{minipage}{0.33\textwidth}
            \includegraphics[width=.9\linewidth]{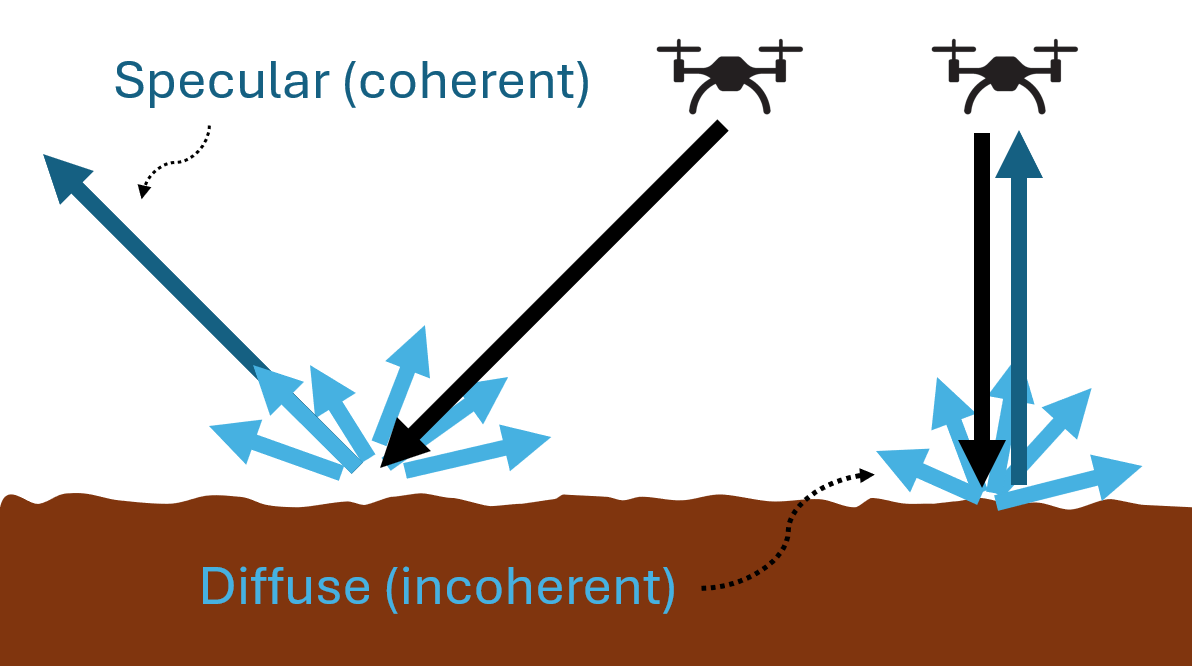}
                      \captionsetup{width=.9\linewidth}

    \caption{Rough soil surface result in specular and diffuse scattering components, while the specular component is dominated in Nadir-looking radars.}
    \label{fig:coherent_incoherent_scattering}
    \end{minipage}
    \begin{minipage}{0.33\textwidth}
        \centering
    \includegraphics[width=.8\linewidth]{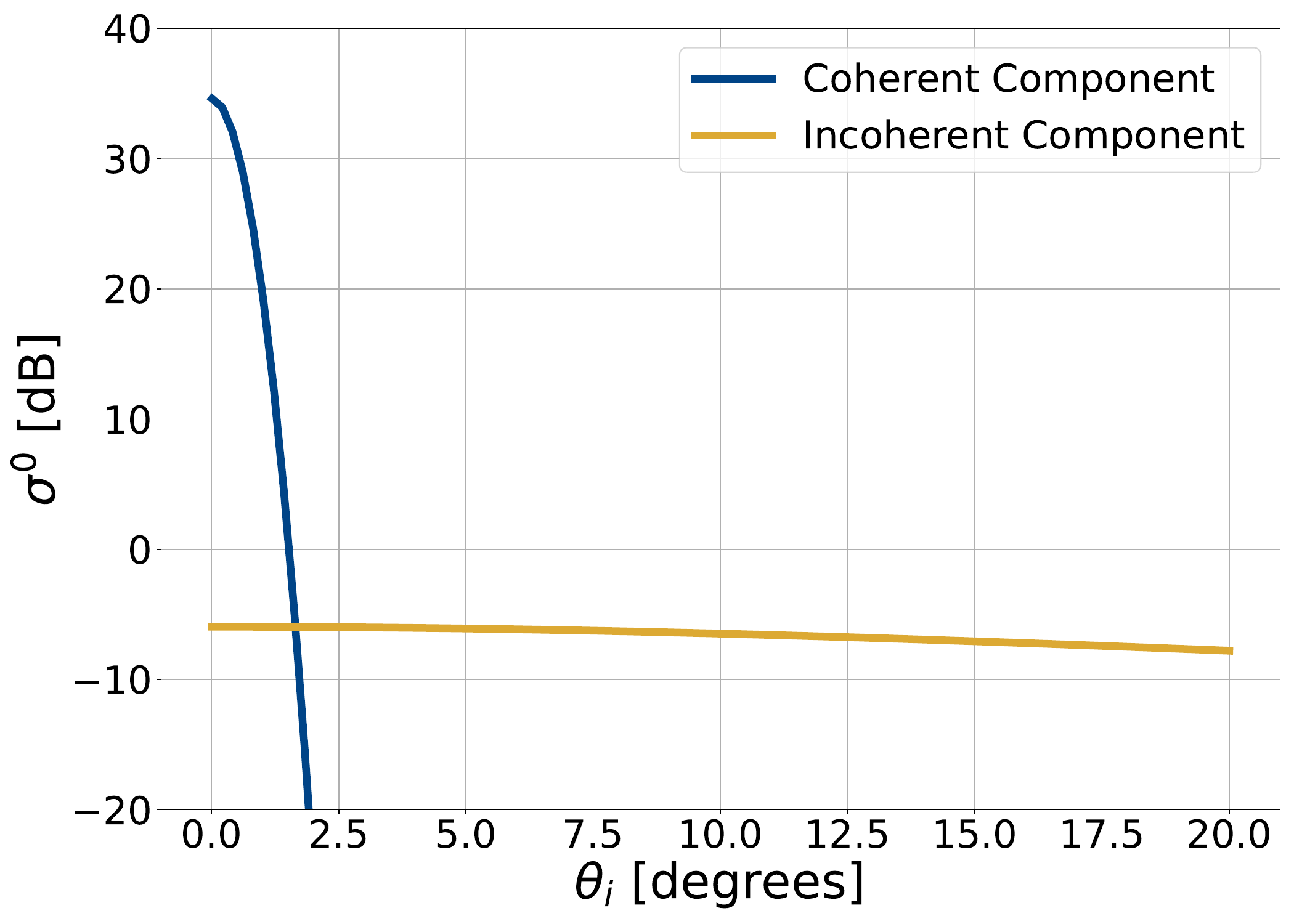}
    \captionsetup{width=.9\linewidth}

    \caption{The coherent ground scattering dominates incoherent component at near nadir angles.}
    \label{fig:anglecoherent}
    \end{minipage}
\end{figure*}
\vspace{.3em} \noindent \textbf{Vegetation Scattering Representation:}
We adopt the approach of \cite{burgin_generalized_2011} to represent the canopy's effect using discrete scatterers. In this approach, the canopy volume is broken down into randomly distributed leaves, branches, and vertically distributed plant trunks. While vegetation canopy has a complicated geometry, it can be approximated into major structural shapes. For example, corn stalks can be approximated as dielectric cylinders, and leaves can be approximated as thin dielectric disks. The far-field scattering behavior of these shapes have analytical solutions derived from Maxwell's Equations and can be encoded into amplitude scattering matrices $\text{S}_\text{cyl}$ and $\text{S}_\text{disk}$. The scattering matrices are a function of radar viewing angle ($\theta_i$, $\phi_i$), scattering angle ($\theta_s$, $\phi_s$), orientation of the scatterer in the elevation and azimuth axes ($\delta$, $\psi$), radius of the scatterer $r$, permittivity of the scatterer $\epsilon_r$, and length of the scatterer $h$:
\begin{equation}
    \text{S}_s = \text{S}_\text{cyl} = \text{S}_\text{cyl}(\theta_i, \theta_s, \phi_i, \phi_s, \psi, \delta, r, h, \epsilon_r)
\end{equation}
\begin{equation}
    \text{S}_l = \text{S}_\text{disk} = \text{S}_\text{disk}(\theta_i, \theta_s, \phi_i, \phi_s, \psi, \delta, r, \epsilon_r)
\end{equation}
The formulas for these scattering matrices are expansive and can be found in \cite{bohren_absorption_2008, durden_modeling_1989}. In the following section, we will see how this scattering representation can be used to calculate total canopy attenuation.


\vspace{.3em} \noindent \textbf{Canopy Attenuation Estimation:} 
As radar signals travel through the canopy, their energy is attenuated by a factor $\Upsilon^2$ due to two-way attenuation. As seen in Figure~\ref{fig:canopy_attenuation}, this attenuation can significantly reduce the amplitude of the ground reflection, so it is necessary to compensate for this effect. The attenuation ($\Upsilon$) through the canopy depends on the total signal path length through the medium and the volume extinction coefficient $\kappa_e$, a measure of the attenuation of the medium per unit length, given as: 
\begin{equation}
\Upsilon = e^{-\frac{\kappa_e h}{\cos \theta_i}}
\end{equation}
, where, $h$ is the depth of the medium and $\theta_i$ is the incidence angle of the wave, and $h/\cos \theta_i$ represents signal path length.
The volume extinction coefficient can be found using the scattering matrices of the medium elements ($\text{S}_s$ and $\text{S}_l$) and their corresponding number densities ($N_s$ for stalks and $N_l$ for leaves):
\begin{equation}
\kappa_e^p = \frac{4 \pi}{k^2} \left( N_{s}\langle \text{Im}[\text{S}_s] \rangle + N_{l}\langle \text{Im}[\text{S}_l] \rangle \right)
\label{eq:volume_extinction_coefficient}
\end{equation}
, where $k = 2\pi f/c$ is the wavenumber and $c$ is the speed of light. As such, \sysname{} canopy model takes into account the cumulative effect of individual stalk and leaf attenuations. Intuitively, the higher the concentration of vegetation stalks or leaves, the more likely that a path through the medium will encounter these elements and be attenuated by them. 
The attenuation effect of individual scatterers can be found by accessing the imaginary component of the scattering matrix $\text{Im}[S]$, averaging its attenuation effect over possible scatterer orientations $\langle \cdot \rangle$, and scaling this effect by density to find the total cumulative attenuation effect: 
\begin{equation} \label{eq:orientation_averaging}
    \langle \text{Im}[\text{S}] \rangle = \iint \text{Im}[\text{S}(\psi, \delta)] p(\psi, \delta) d\psi d\delta
\end{equation}
We model stalks to only be oriented vertically, but leaves can be oriented across a range of orientations $(\psi, \delta)$, and the PDF of this orientation probability distribution $p(\psi, \delta)$ encodes the range of orientations. We use a uniform probability distribution for soybean leaves as they are randomly oriented.

\vspace{.3em} \noindent \textbf{Ground Reflection Model:}
Next, we model the ground radar cross-section (RCS), which carries soil permittivity information, using a rough-surface dielectric scattering model. Scattering from a rough surface can be decomposed into two components (shown in Figure \ref{fig:coherent_incoherent_scattering}): a specular (coherent) component that radiates from the surface at an elevation angle equal to the incidence angle and a diffuse (incoherent) component that radiates in all directions and increases in power with surface roughness. 
Prior soil moisture retrieval algorithms for side-looking radars have relied on incoherent component models and have discarded the coherent component which is insignificant in the side-looking backscatter direction. However, the coherent component is dominant at nadir look angles, so we extend an existing incoherent soil backscatter model \cite{tabatabaeenejad_bistatic_2006} to capture this effect. \sysname{} models the coherent component using the Kirchhoff Approximation \cite{fung_coherent_1983} given as:
\begin{equation}
\sigma_\text{coh}(\theta_i, f) = A \frac{\Gamma}{\beta_c^2} \exp\left(-\frac{16\pi^2s^2f^2}{c^2}\right) \exp\left(\frac{\theta_i^2}{\beta_c^2}\right),
\end{equation}
 where $\Gamma$ is the Fresnel reflectivity at normal incidence, $\beta_c$ is a scattering beamwidth tuning parameter, $s$ is the surface roughness height, and $A$ is the surface area illuminated by the radar pulse. The Fresnel reflectivity is a function of the soil permittivity $\epsilon_s$: 
\begin{equation}
    \Gamma = \left| \frac{1 - \sqrt{\epsilon_s}}{1 + \sqrt{\epsilon_s}} \right|^2
\end{equation}
Through the Topps equation \cite{topp_electromagnetic_1980}, soil permittivity is linked to soil moisture, so variations in soil moisture will produce variations in the coherent RCS contribution. 



\vspace{.3em} \noindent \textbf{Extending Ground Reflection Model to Radar Beam:}
The radar illuminates a wide footprint on the ground determined by its half-power beamwidth. However, most of the received ground-reflected power originates from a much smaller region near nadir, where coherent specular reflection dominates. To approximate the total simulated RCS, we therefore consider only the effective ring area $A$ that captures this coherent reflection energy. This area expands or contracts with changes in drone altitude $R$ because the incidence angle influences the extent of coherent reflection. However, the corresponding effective coherent beamwidth $\theta_\text{e}$ remains essentially invariant to altitude, making it a more stable descriptor of the radar’s spatial response. Given the radar altitude $R$ and coherent beamwidth $\theta_\text{e}$, the effective reflective area can be approximated as: 
\begin{equation}
    A = \pi \left(R \tan(\theta_\text{e}/2) \right)^2
\label{eq:effective_reflective_area}
\end{equation}
Figure~\ref{fig:anglecoherent} shows the simulated ground RCS as a function of incidence angle, including both coherent and incoherent scattering components. The results indicate that coherent scattering dominates the total RCS at small incidence angles, particularly below 2°, where the specular reflection from the ground surface is strongest.



\vspace{.3em} \noindent \textbf{Putting It All Together:}
Finally, we combine the canopy attenuation with ground RCS by multiplying the two-way transmittance of the canopy: 
\begin{equation}
\sigma_s(f) = \Upsilon^2(\theta_e, f) \sigma(\theta_e, f),
\label{eq:sigma_s}
\end{equation}
where $\theta_e$ is the effective radar beamwidth with dominant coherent component. 




\subsection{Radar Frequency Response Processing \& RCS Computation}
As described in the previous section, \sysname{}’s radiative transfer forward model expresses the canopy-attenuated ground return as a frequency-dependent radar cross-section (RCS). To be compatible with this model, we therefore need to isolate the ground-reflection component from the measured radar waveform and represent it in the same RCS format. However, typical wideband radars employ pulse-shaped excitations (e.g., Ricker wavelets) and acquire time-domain signals. \change{Simply applying an FFT to the time-domain signal produces a frequency-domain representation that aggregates components from direct antenna coupling, canopy returns, and other environmental scatterers, and therefore cannot isolate the ground response.
} Furthermore, the transmit waveform itself exhibits a non-flat spectrum, which can obscure the true channel frequency response associated with the ground reflection and the underlying soil permittivity. \change{This non-flat spectrum in the transmit waveform is a technical challenge to wideband RCS estimation, as variations of energy across the pulse spectrum can distort the RCS across the frequency band.} 

To address these challenges, we develop a signal-processing pipeline that (1) first extracts the ground return in the time domain using range gating and converts the measurement into the frequency domain, and (2) maps the resulting channel frequency response into a calibrated RCS spectrum by removing waveform-dependent spectral shaping and makes it compatible with our forward model. 



\begin{figure}
    \centering
    \includegraphics[width=1.0\linewidth]{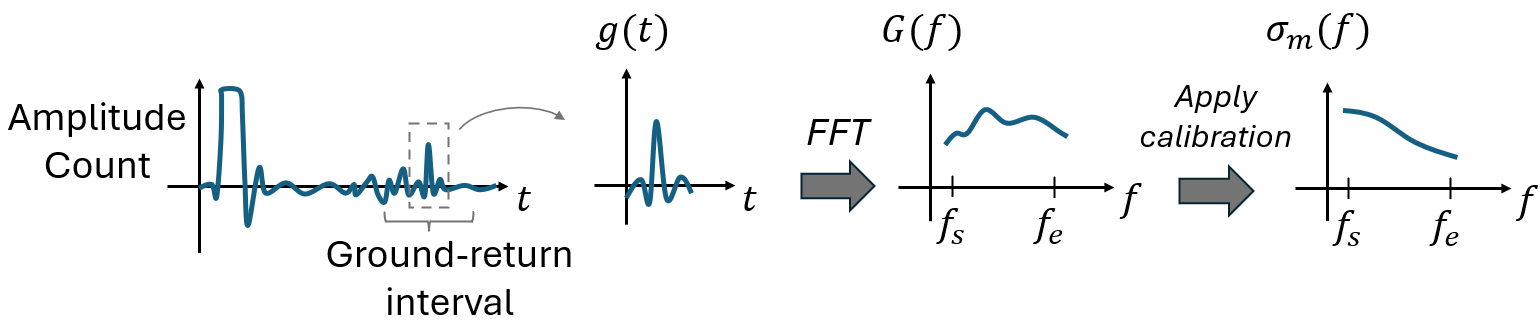}
    \caption{\sysname{} pipeline from raw radar measurement to ground RCS estimation}
    \label{fig:rcs_conversion}
\end{figure}

\vspace{.3em} \noindent \textbf{Ground-Return Isolation and Frequency Response Estimation:} The radar measurement provides a time-domain waveform, often referred to as an A-scan (Figure ~\ref{fig:rcs_conversion}), in which a pulse at time $t$ corresponds to a reflection from a target at the range implied by that time-of-flight. Because our transmit pulse is wideband, the measurement has sufficient temporal resolution to distinguish closely spaced scatterers along the range dimension. To isolate the ground contribution, we determine the expected ground-return interval using the drone altitude provided by GPS or an onboard altimeter. \change{Only a rough estimate of the drone altitude is necessary ($\pm$ 10 cm), as this altitude will only be used to center a coarse search interval.} Within this search interval, we locate the dominant peak in the signal envelope and extract a short window centered on this peak, with the window length chosen based on the known pulse duration. This yields a time-domain segment that predominantly contains the ground reflection while suppressing earlier canopy returns and noise outside the ground-return region. \change{This ground reflection segment, however, contains some clutter from scatterers such as leaves near the ground, but the ground reflection dominates the clutter, as the ground scatters coherently and in the direction of the radar, while angled leaves will scatter incoherently and in many directions.} We then convert the ground reflection segment $g(t)$ into a reflection spectrum $G(f)$ using the FFT operation. The square of this reflection spectrum is proportional to the energy received by the radar at individual frequencies.

\vspace{.3em} \noindent \textbf{Ground-return RCS Calculation:} The ground-return frequency response $G(f)$ computed in the previous step is still a system–path response as it contains the imprint of the antenna patterns, propagation losses, waveform spectral shaping, receiver gain, and other hardware effects. In principle, one could invert these factors using the radar equation by explicitly applying known antenna gain patterns, transmit power, and receiver noise figure to recover RCS analytically. However, this approach requires a complete and accurate parametrization of all radar hardware terms, which is often not fully available. Instead, we adopt a one-time calibration approach that indirectly estimates the required scaling by measuring the frequency response of a reference target with known analytical RCS. We use the method of Melebari et al. \cite{melebari_absolute_2024} by introducing a frequency-dependent calibration factor $C(f)$ that can be used to obtain $\sigma_m(f)$ from $G(f)$:
\begin{equation}
\sigma_m(f) = \frac{f^2 R^4 |G(f)|^2}{C(f)}
\label{eq:measured_rcs}
\end{equation}
This calibration factor is found by measuring the frequency response $G_r(f)$ of a reference target (e.g. a metal plate) with known RCS $\sigma_r$ at a known range $r_r$:
\begin{equation}
C(f) = \frac{f^2 r_r^4 |G_r(f)|^2}{\sigma_r}
\label{eq:calibration_factor}
\end{equation}
Once $C(f)$ is determined, any isolated ground-return frequency response $G(f)$ can be mapped to a calibrated ground-return RCS spectrum $\sigma_m(f)$. This approach avoid reliance on detailed hardware modeling and takes care of waveform spectral shaping, while still yielding physically interpretable scattering quantities compatible with our forward model. 

\subsection{Joint Canopy-Soil Moisture Retrieval}
During retrieval, \sysname{} jointly estimates soil and canopy permittivities ($\epsilon_s$ and $\epsilon_c$) by inverting the proposed RCS model:
\begin{equation} \label{eq:retrieval_equation}
\{\epsilon_s, \epsilon_c\} = \arg \min_{\hat \epsilon_s, \hat \epsilon_c} \sum_i^M \left( \sigma_s(f_i) - \sigma_m(f_i) \right)^2
\end{equation}
where $\sigma_s(f_i)$ represents the simulated ground RCS for frequency $f_i$ and $\sigma_m(f_i)$ is the measured ground RCS from the radar data, and $M$ represents the number of frequency components extracted from time-domain radar measurements. The radiative transfer model depends on several physical parameters, including soil surface roughness and canopy structure (Figure~\ref{fig:backscatter_sidelooking_vs_nadirlooking}).

\vspace{.3em} \noindent \change{\textbf{Soil Roughness Estimation:} }\change{Soil roughness height is empirically measured from bare-soil data collected at the beginning of the growing season. Collecting this calibration data involves capturing a radar scan over a bare region of the field and recording the soil moisture of the region using a soil moisture probe or oven-dry technique. Using the recorded soil moisture and the radar scan, the soil roughness height is calibrated by finding the roughness height that minimizes the soil moisture retrieval error. Soil roughness height is assumed constant throughout the growing season, as surface roughness under canopy varies minimally over time. In the case that a soil-disturbing machine such as a cultivator or rotary hoe needs to be used on the field after the canopy is grown, this empirical calibration may be performed at a later time using a bare region of the same field.}

Canopy structural parameters, including height and density, are estimated from LiDAR data as described in the following section.

\begin{figure}
    \centering

    \includegraphics[width=.99\linewidth]{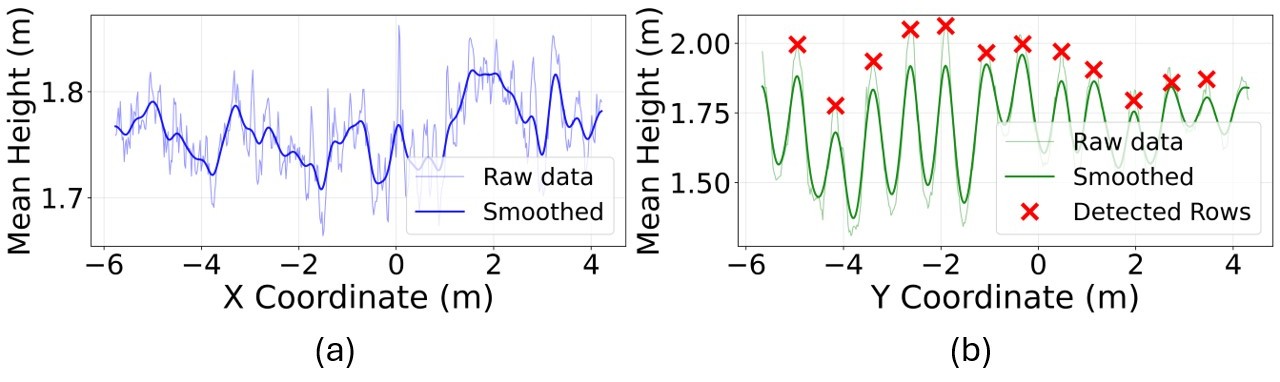}

    \begin{subfigure}{\linewidth}
        \centering
        \captionsetup{labelformat=parens} 
        \renewcommand{\thesubfigure}{c}   
        \includegraphics[width=\linewidth]{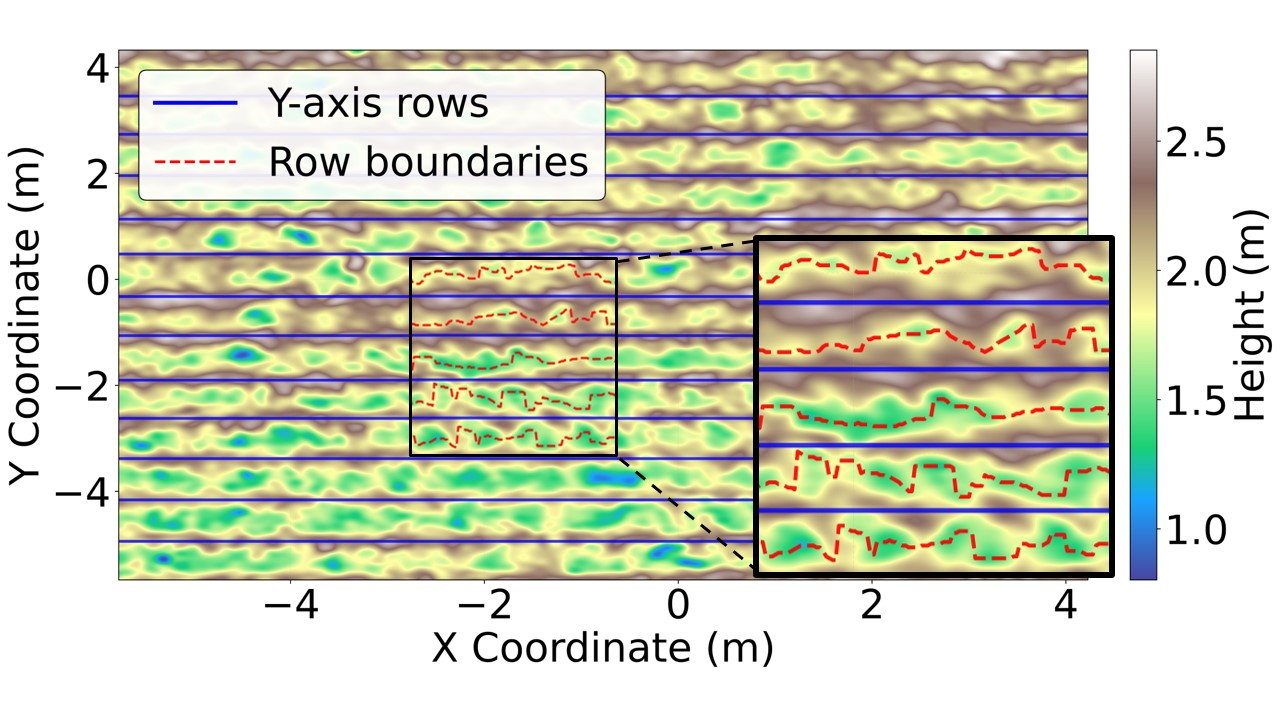} 
        \caption{}
    \end{subfigure}

    \caption{Example height profile along (a) X axis without regular row structure, and (b) Y axis with periodic peaks detected as rows; (c) Example of Row lines detected and visualized inter-row boundary delineation}
    \label{fig:chm_height_profiles}
\end{figure}

\subsection{Lidar-Based Canopy Structure Extraction}

To estimate canopy-induced attenuation in radar returns, \sysname{} leverages LiDAR data to estimate canopy structural parameters such as height, spatial density, leaf size, and leaf density. These canopy structural parameters serve as input to \sysname{} radiative transfer model for jointly estimating soil and canopy permittivity. However, the key challenge is that LiDAR scanning of vegetation experiences significant measurement noise due to factors such as wind-induced canopy motion and errors in the laser returns. The resulting point clouds exhibit noisy and discontinuous plant and foliage structures. To address this challenge, \sysname{} first extracts canopy height and row-wise plant density directly from UAV-mounted LiDAR point clouds by leveraging the row structure typical in crop fields.It then uses crop-specific allometric relationships derived from field measurements together with LiDAR-derived structural features to estimate leaf size and leaf density. This produces a vegetation description that is consistent with both the physical crop geometry captured by LiDAR and the expected plant morphology predicted by agronomic allometry.

\begin{figure}
    \centering

    \begin{subfigure}{\linewidth}
        \centering
        \includegraphics[width=.99\linewidth, height=1.74cm]{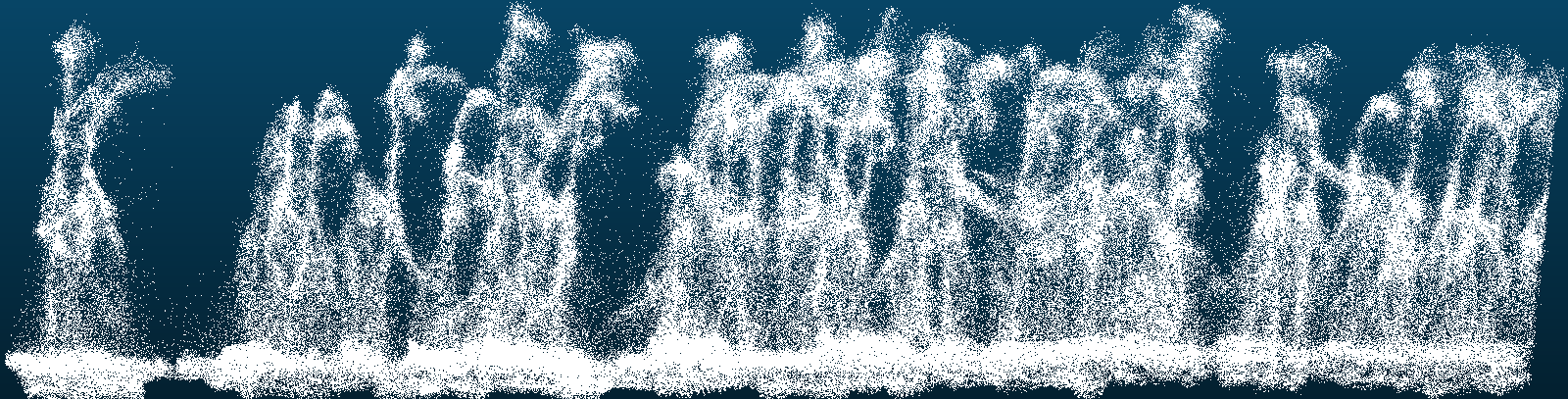}
        \caption{}
    \end{subfigure}

    \begin{subfigure}{\linewidth}
        \centering
        \includegraphics[width=.99\linewidth]{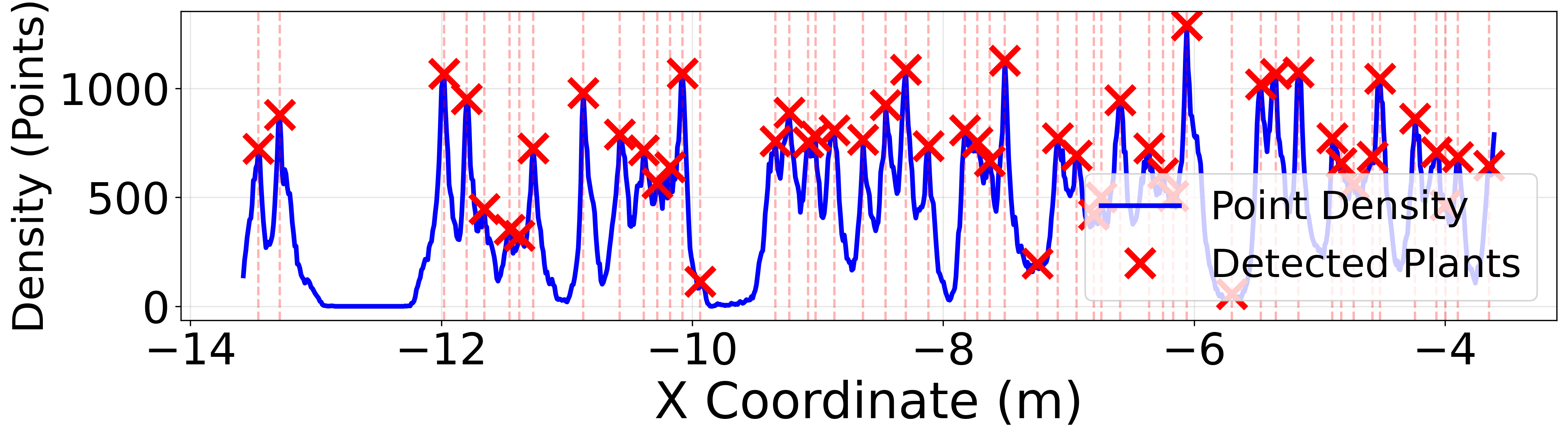}
        \caption{}
    \end{subfigure}

    \begin{subfigure}{\linewidth}
        \centering
        \includegraphics[width=.99\linewidth]{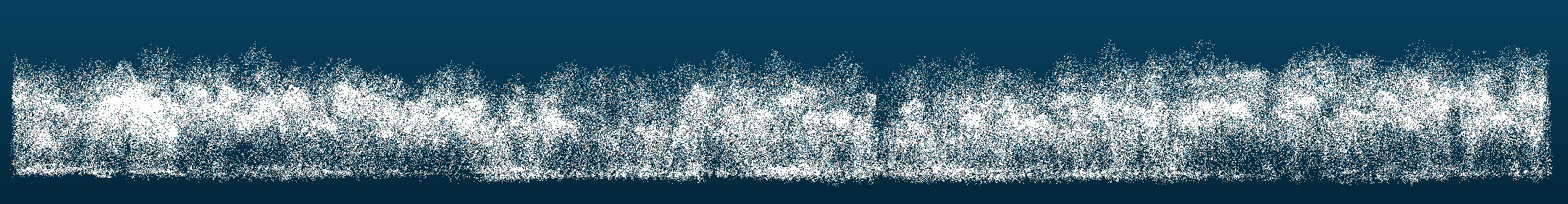}
        \caption{}
    \end{subfigure}

    \begin{subfigure}{\linewidth}
        \centering
        \includegraphics[width=.99\linewidth]{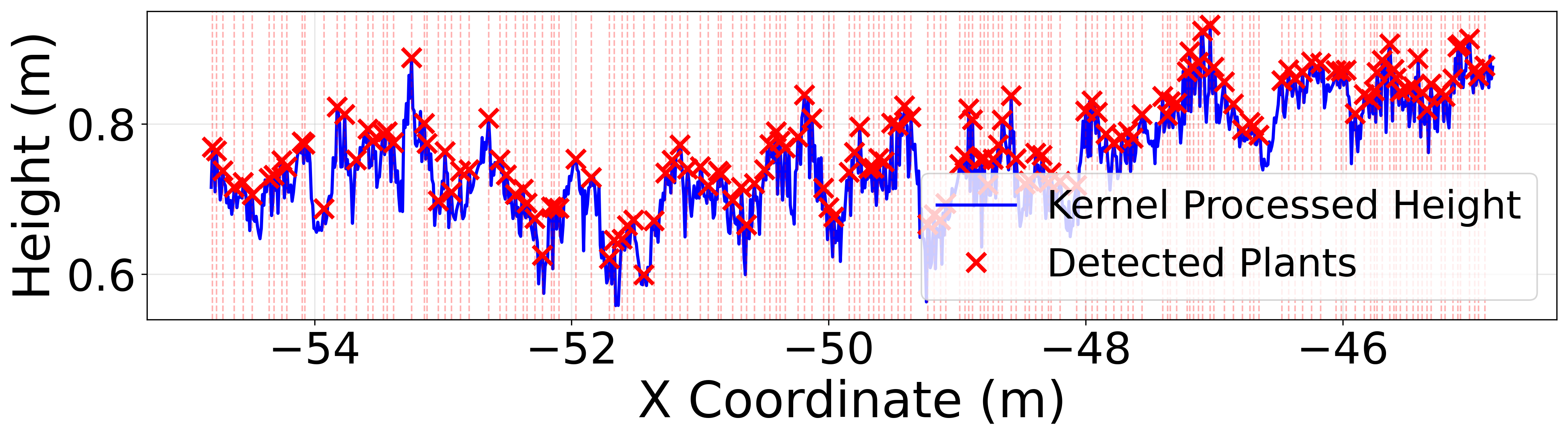}
        \caption{}
    \end{subfigure}

    \caption{(a) Lidar data for a single corn canopy row and (b) its corresponding density map along the row; (c) Lidar data of a single soybean row and (d) corresponding Kernel-processed height map along the row}
    \label{fig:corn_soybean_map}
\end{figure}

\vspace{.3em} \noindent \textbf{Row-Aware Canopy Segmentation:} 
\sysname{} first normalizes LiDAR point clouds relative to the local ground surface, mapping the ground to $z=0$ and aligning crop rows parallel to either the x- or y-axis. The field is then divided into 10~m~$\times$~10~m tiles. Within each tile, \sysname{} generates a canopy height model (CHM) \cite{Fischer2024_CHM} at 2~cm resolution, followed by Gaussian smoothing ($\sigma=3$~pixels) to suppress small-scale noise. The CHM represents the height of the vegetation canopy above the ground by recording the maximum elevation within each grid cell. The row direction is inferred by projecting the CHM onto both axes and selecting the one with stronger periodicity in its height profile. Periodicity is quantified by $\text{score}=1/(1+\mathrm{CV})$, where $\mathrm{CV}$ is the coefficient of variation of distances between adjacent height peaks. As shown in Fig.~\ref{fig:chm_height_profiles}, once the dominant orientation is determined, crop row centerlines are extracted by locating local minima between neighboring height peaks on the smoothed CHM profile. Examples resulting segmented rows are illustrated in Fig.~\ref{fig:corn_soybean_map}(a)(c).


\vspace{.3em} \noindent \textbf{Canopy Height and Density Estimation:} After each row of crops are segmented, \sysname{} estimates canopy structure of each row using two complementary strategies. For tall crops with distinct stems such as corn, stem locations produce locally high point concentrations in the LiDAR data even when the upper canopy appears merged (shown in Fig.~\ref{fig:corn_soybean_map}(a)), so \sysname{} computes a height-weighted point-density profile perpendicular to each row and detects peaks as plant centers (shown in Fig.~\ref{fig:corn_soybean_map}(b)); peak count per unit area yields plant density. For low-stature crops such as soybean, where stems visually merge (shown in Fig.~\ref{fig:corn_soybean_map}(c)), \sysname{} instead applies a 3D rectangular kernel on the smoothed CHM along the row direction. The kernel summarizes local canopy height to produce a height profile along the row direction (Fig.~\ref{fig:corn_soybean_map}(d)). Peaks in this profile correspond to plant centers. In both cases, plant regions are defined using midpoints between adjacent detected plant centers. 

\vspace{.3em} \noindent \textbf{Leaf Density Estimation:}
\sysname{}'s radar attenuation model requires estimates of both leaf density and average single-leaf area. However, direct extraction of leaf surfaces from the LiDAR is unreliable because sampling noise and leaf motion fragment foliage patterns \cite{Haluboka2021, Wang2020Review}. Instead, \sysname{} estimates the leaf area index (LAI), which defines the total leaf area per unit area (m$^2$/m$^2$), from the point cloud and derives leaf density using crop-specific mean leaf area from empirical studies.
To estimate LAI, \sysname{} adopts a simplified voxel-based Beer–Lambert inversion commonly used in LiDAR canopy studies \cite{Beland2014Voxel, Beland2021, Haluboka2021, Nguyen2022}. Each field tile is divided into eight vertical layers, and LiDAR point returns within each layer are used to approximate canopy gap probability and attenuation. LAI is then estimated as
\begin{equation}
\text{LAI} = -\frac{1}{G} \sum_k 
\frac{\ln P_\text{gap}(z_{k+1}) - \ln P_\text{gap}(z_k)}{v_z},
\end{equation}
where $P_\text{gap}(z_k)$ represents the probability that a LiDAR pulse passes through the canopy without interception, $v_z$ is the vertical spacing between layers, and $G$ is the leaf projection coefficient (typically $\approx 0.5$ for randomly oriented leaves). This approach provides a lightweight and physically interpretable estimate of canopy structure suitable for UAV LiDAR data. Finally, leaf density is obtained as $D_\text{leaf} = \text{LAI}/A_\text{leaf}$, where $A_\text{leaf}$ is the average single-leaf area obtained from field measurements.

%% file: sections/4-implementation.tex
\section{Implementation}

\begin{figure}[t]
    \centering
    \begin{minipage}{0.49\linewidth}
        \centering
        \includegraphics[width=\linewidth]{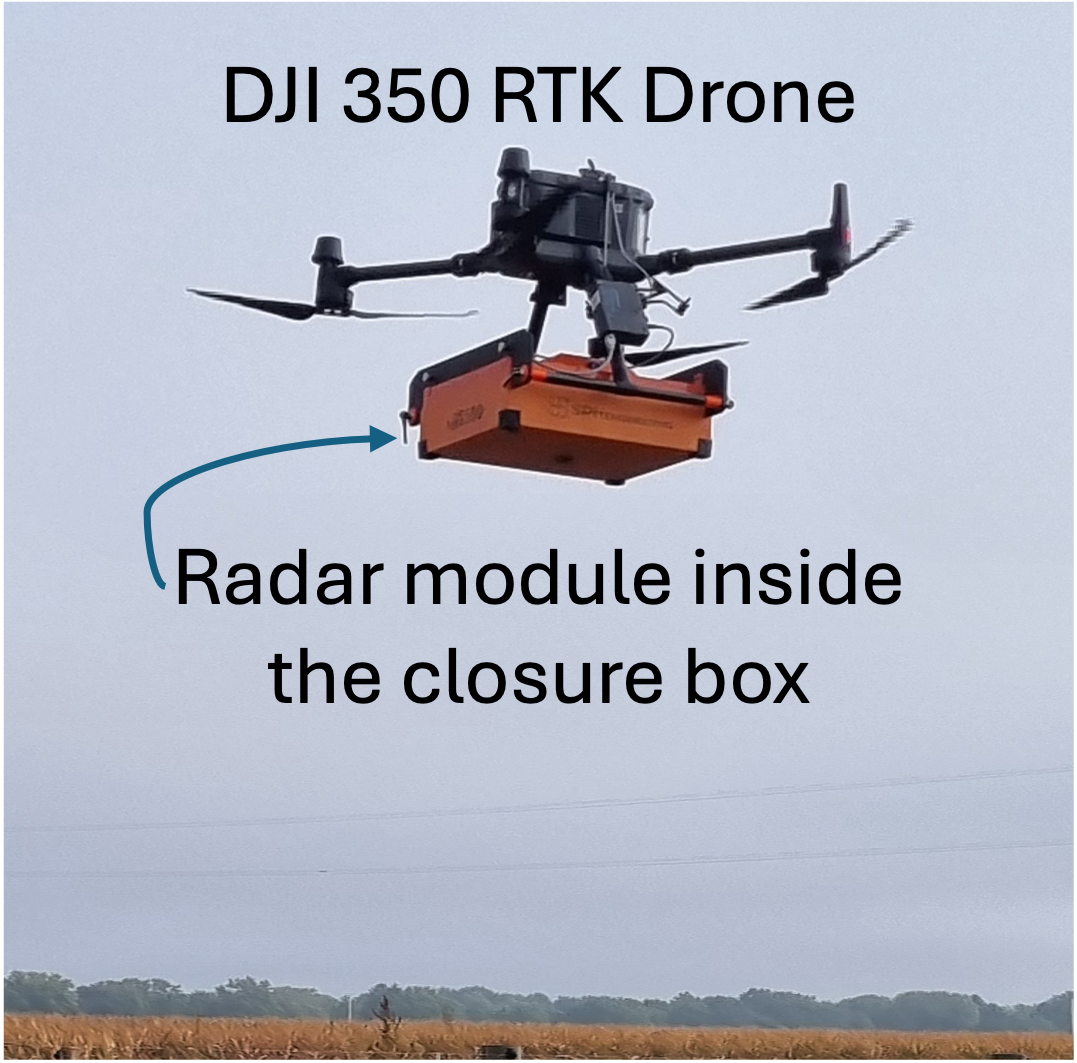}
        \caption{Our UAV prototype}
        \label{fig:setup}
    \end{minipage}\hfill
    \begin{minipage}{0.49\linewidth}
        \centering
        \includegraphics[width=\linewidth]{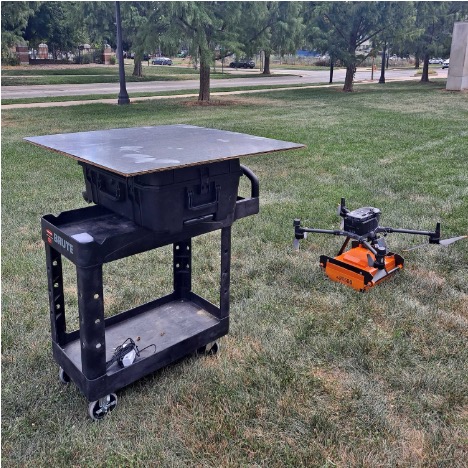}
        \caption{Calibration setup}
        \label{fig:drone_and_metal_plate}
    \end{minipage}
        
\end{figure}

\noindent \textbf{Aerial Platform.} Figure \ref{fig:setup} illustrates the \sysname{} prototype, which comprises three main components: (1) A Zond Aero 500 NG ground-penetrating radar transmits Ricker pulses across a 200–900 MHz bandwidth, providing a balance between canopy penetration and resolution while remaining compliant with FCC GPR regulations. The received electric field is directly sampled using a high-speed analog-to-digital converter (ADC) operating at approximately 14 GS/s. These time-domain signals are referred to as "A-scans" by the GPR community. The radar employs wideband bowtie antennas with an azimuth half-power beamwidth of 60 degrees, offering broad coverage and a stable phase center across the operating bandwidth. (2) The radar is mounted on a DJI Matrice 350 RTK drone and integrated with the flight control system for GPS-based georeferencing. (3) The platform is also equipped with a DJI L2 LiDAR sensor for canopy structure characterization. \change{While the current platform cost over \$30,000 USD, this is a one-time up-front cost that does not scale with field size, an advantage of an infrastructure-free approach. Moreover, the time and labor cost to operate the platform is less than prior work that requires burying and maintaining infrastructure.}

\begin{figure}
    \centering
    
    \begin{subfigure}{0.49\linewidth}
        \centering
        \includegraphics[width=\linewidth]{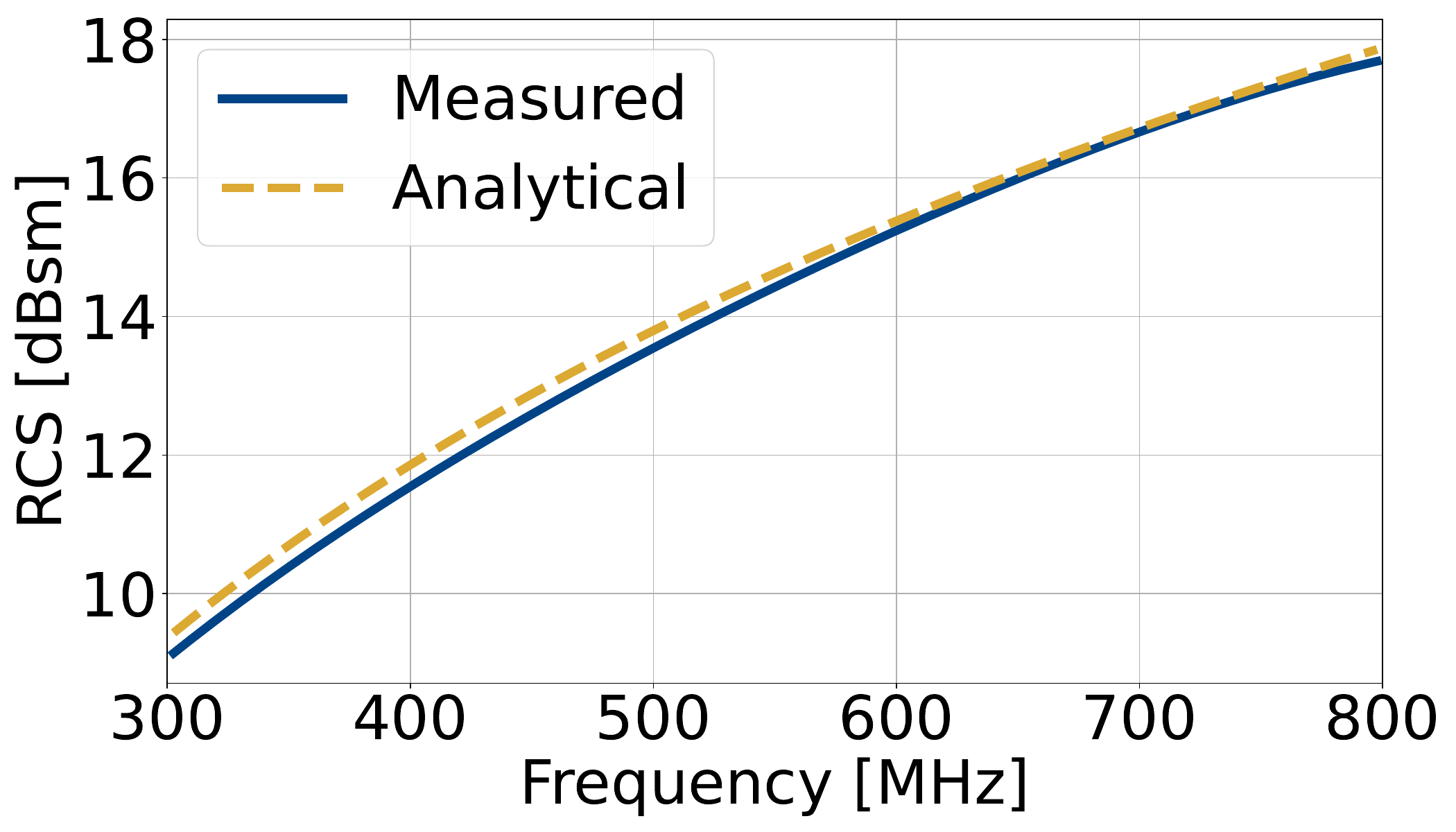}
        \caption{}
        \label{fig:rcs_stability}
    \end{subfigure}
    \hfill
    \begin{subfigure}{0.49\linewidth}
        \centering
        \includegraphics[width=\linewidth]{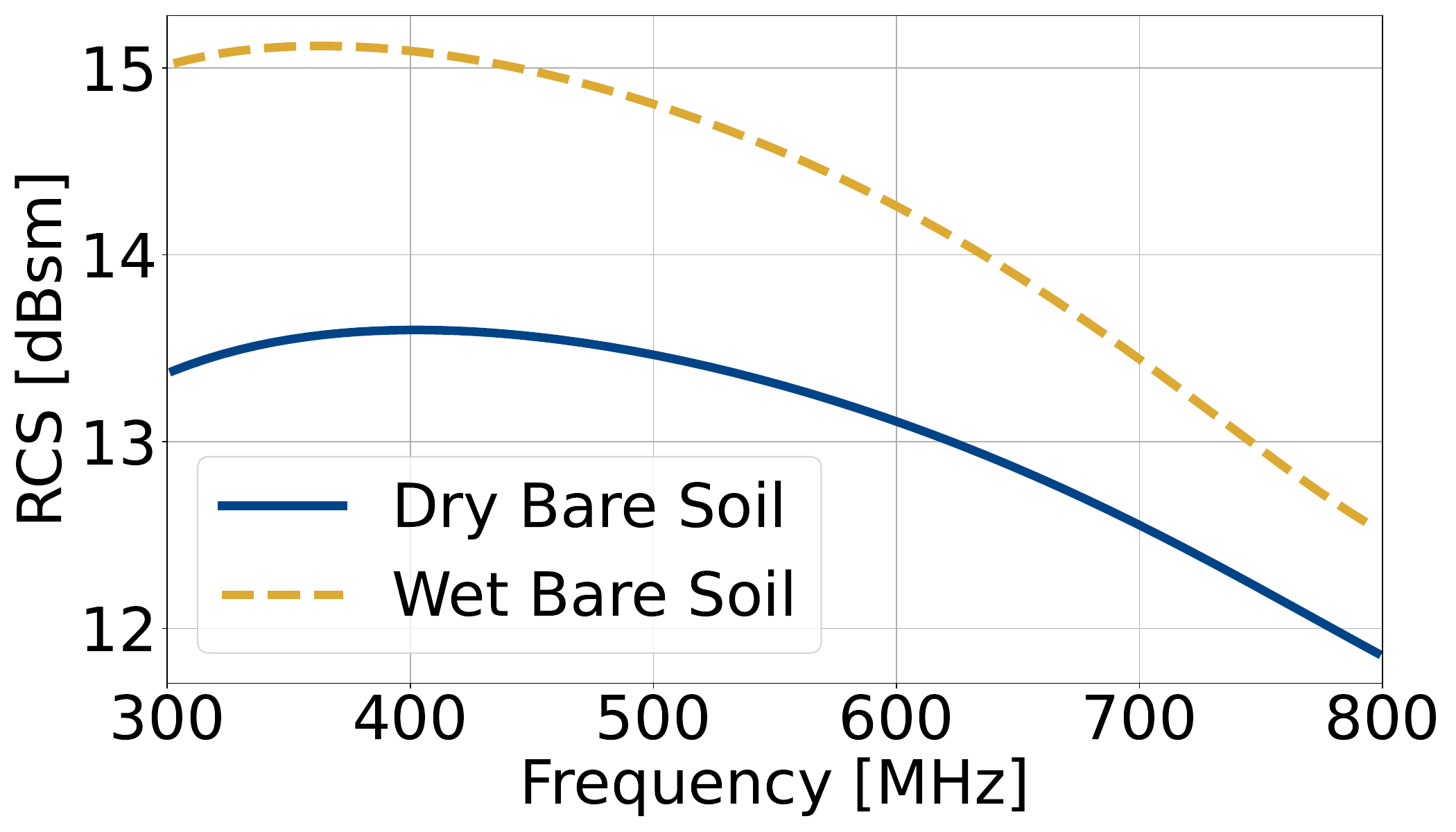}
        \caption{}
        \label{fig:rcs_soils}
    \end{subfigure}
    
    \captionsetup{width=\linewidth}
    \caption{Validation of (a) metal plate RCS with analytical model and (b) calibrated bare-soil RCS under soil moisture conditions (dry is 11\% VWC and wet is 21\% VWC).}
    \label{fig:calibValid}
\end{figure}


\vspace{.3em} \noindent \textbf{Retrieval Algorithm.} 
We implemented the retrieval algorithm in Python, building upon the canopy radiative transfer framework described in \cite{burgin_generalized_2011}. In our forward model, soybean leaves were represented as dielectric disks with diameters equal to the measured leaf lengths. For corn, whose leaves are more elongated, we approximated each leaf as a concatenation of dielectric disks with diameters equal to the leaf width, yielding an effective leaf geometry that preserves the overall length and aspect ratio. For the inverse retrieval, we jointly estimated canopy and soil moisture by solving the optimization problem of Equation~\ref{eq:retrieval_equation}. We used a grid search optimizer to search a wide range of complex permittivity values for both layers, discretizing the joint canopy-moisture search space into 500 evenly-spaced values across both dimensions.


\subsection{One-Time RCS Calibration} 
\sysname{} leverages a one-time calibration approach to scale RCS measurements based on the radar hardware parameters such as antenna pattern and gain, or waveform spectral shaping. This is done using a reference target with known analytical RCS. Inspired by the calibration procedure proposed in \cite{melebari_absolute_2024}, we capture the radar return on top of a square metal plate of length 90 cm (shown in Figure \ref{fig:drone_and_metal_plate}). The metal plate was elevated 1 m above the ground to better isolate the plate reflection from ground reflection in the time range domain. We capture the plate reflection at 7 different altitudes between 6-9 m and estimated the mean $C(f)$ calibration factor using the PEC plate RCS analytical expression of
\begin{equation}
    \sigma_r(f) = 4\pi l^4 / \lambda^2
\end{equation}
where $\lambda$ represents the wavelength and $l$ is the one-side dimension of the metal plate. \change{This calibration setup is practical and inexpensive, requiring only a radar calibration target such as a metal plate.} 

It is worth noting that this calibration only needs to be performed once for the radar and is independent of location or individual data collection trials. To verify the stability of the calibration factor, we applied the factor obtained from one experiment to estimate the radar cross section (RCS) of a metal plate in a different field, with the drone flying at a different altitude, using Equation~\ref{eq:measured_rcs}. The results, shown in Figure~\ref{fig:calibValid}(a), demonstrate agreement within 1 dBsm between the analytical and measured RCS values across fields. This confirms that the calibration factor remains stable over time and provides accuracy within 1 dB for estimating the RCS of far-field scatterers in the nadir direction.

We further evaluated the impact of this calibration factor on soil sensing using a bare-soil field under two distinct moisture conditions. The calibration effectively removes hardware-related effects while preserving the intrinsic non-linear frequency response. Moreover, when comparing wet and dry soil in Figure~\ref{fig:calibValid}(b), we observe a higher RCS for wet soil, which aligns with analytical expectations due to its higher permittivity.

%% file: sections/5-evaluation.tex
\section{Evaluation}


\begin{figure}
    \centering
    \includegraphics[width=.49\linewidth]{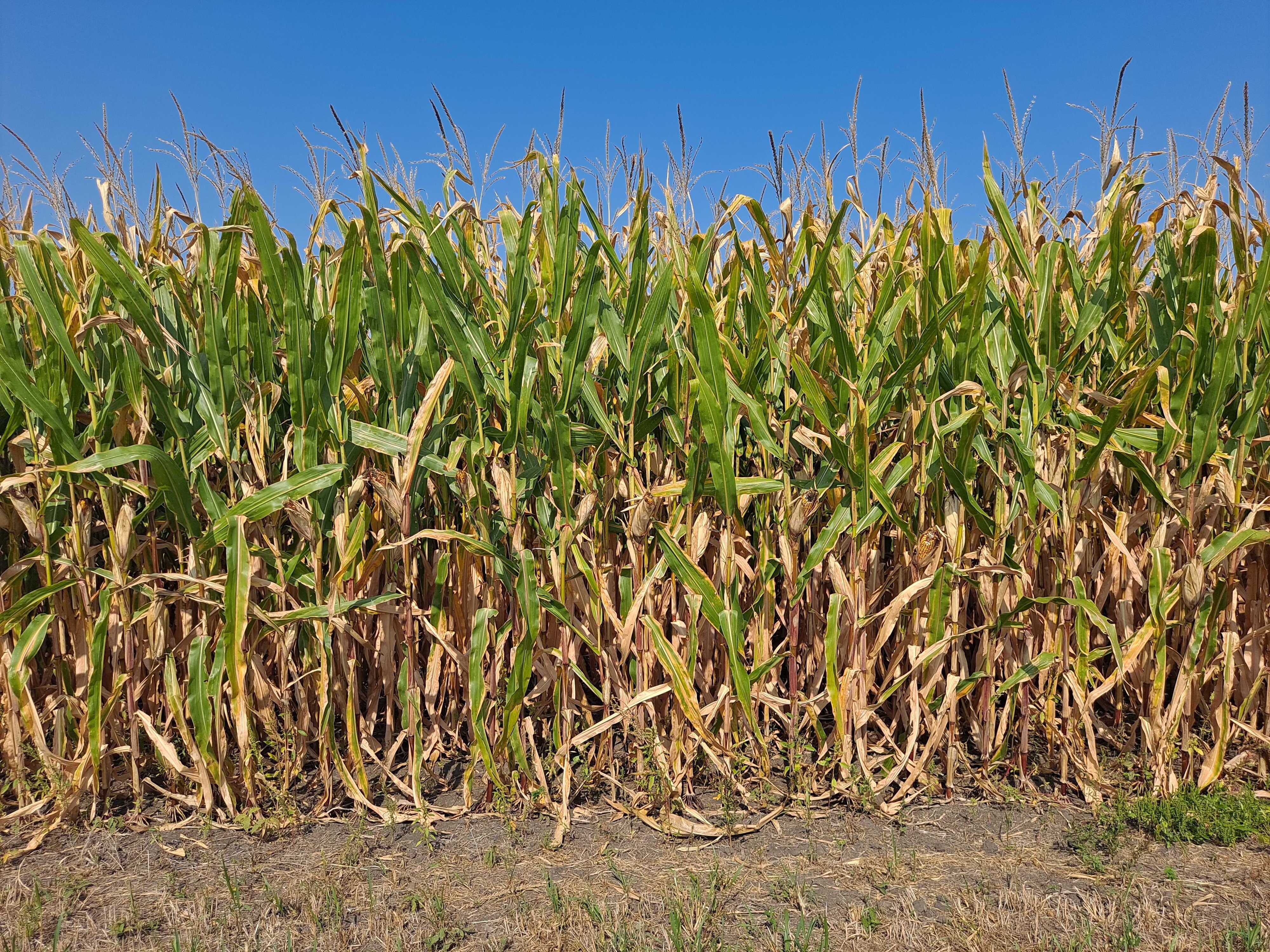}
    \includegraphics[width=.49\linewidth]{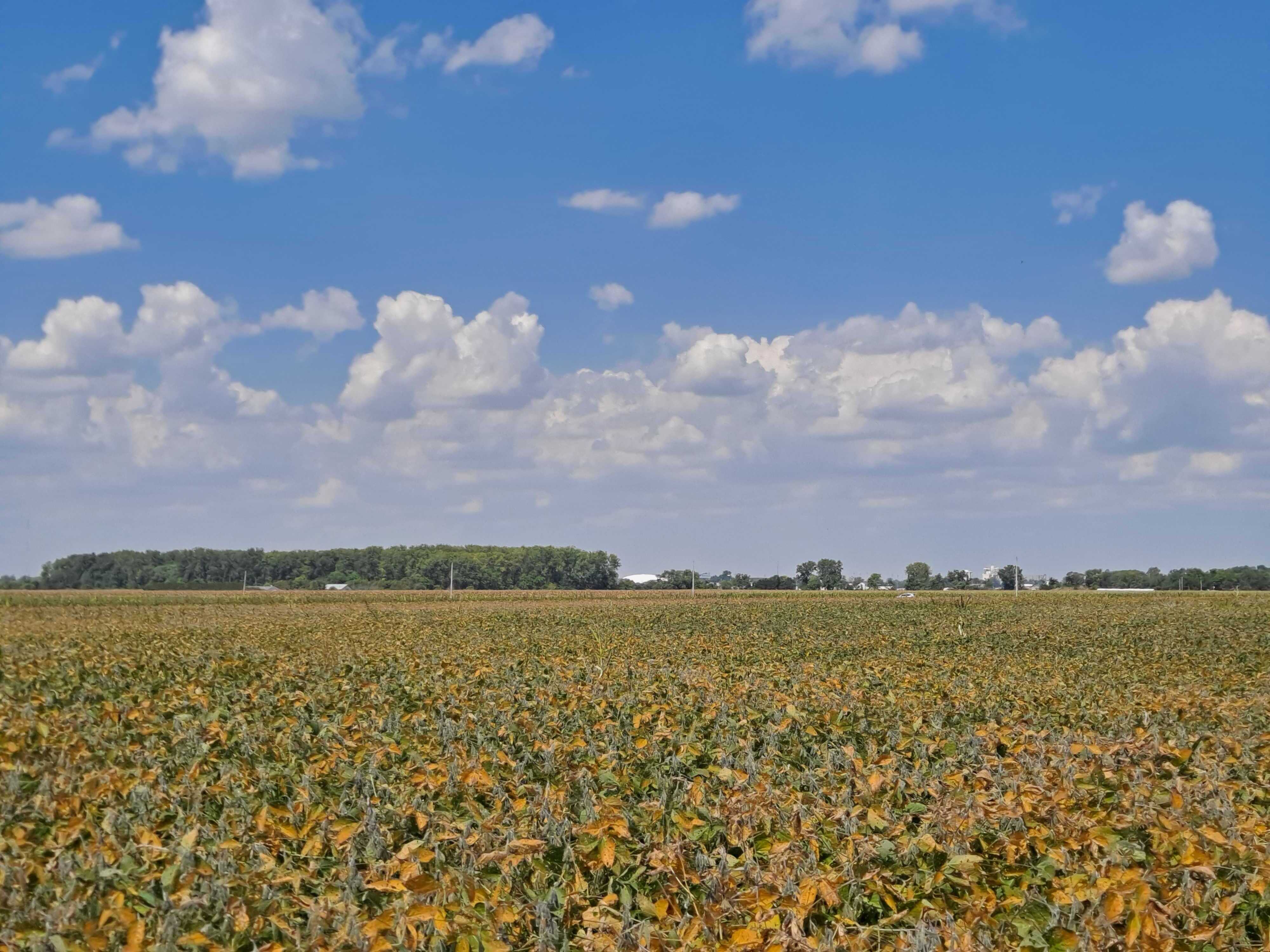}
    \includegraphics[width=.49\linewidth]{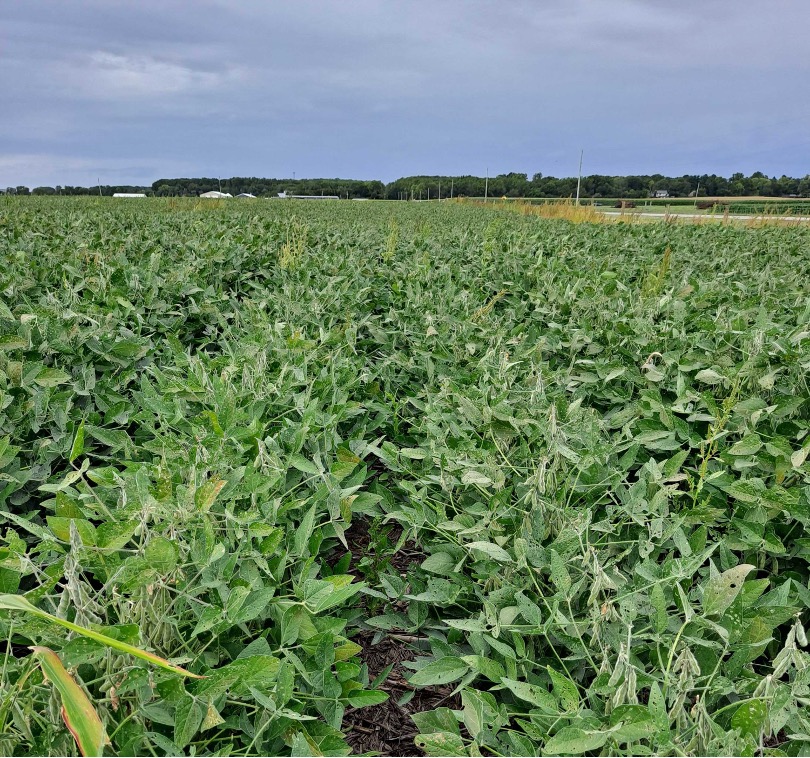}
    \includegraphics[width=.49\linewidth]{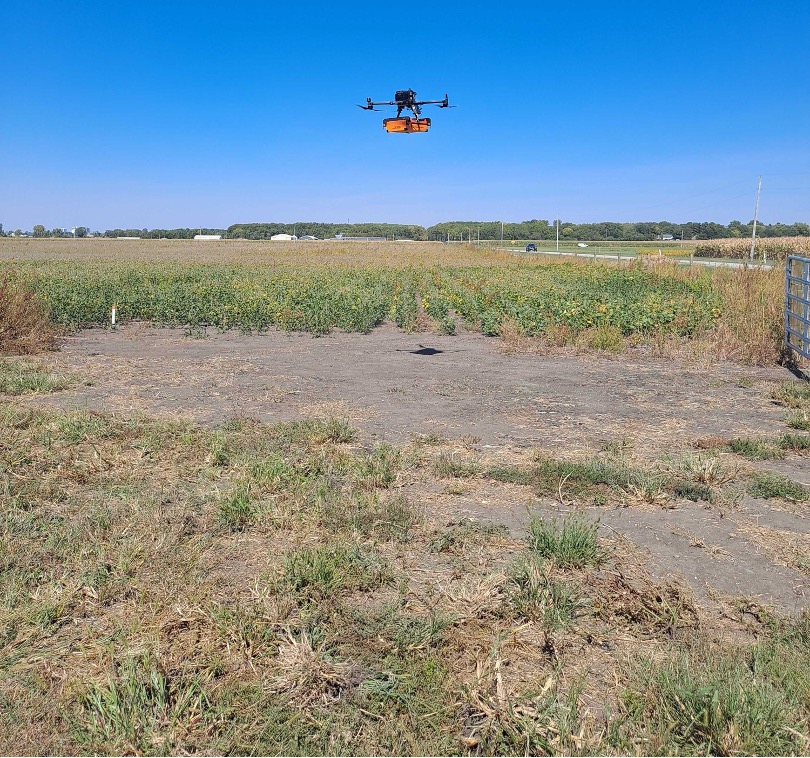}
    \caption{Snapshots of our experimental sites with different canopy coverage, spanning over 9 acres.}
    \label{fig:canopy_pictures}
\end{figure}


\subsection{Experimental Setup}

\begin{figure}[t]  
\centering
\includegraphics[width=0.32\linewidth]{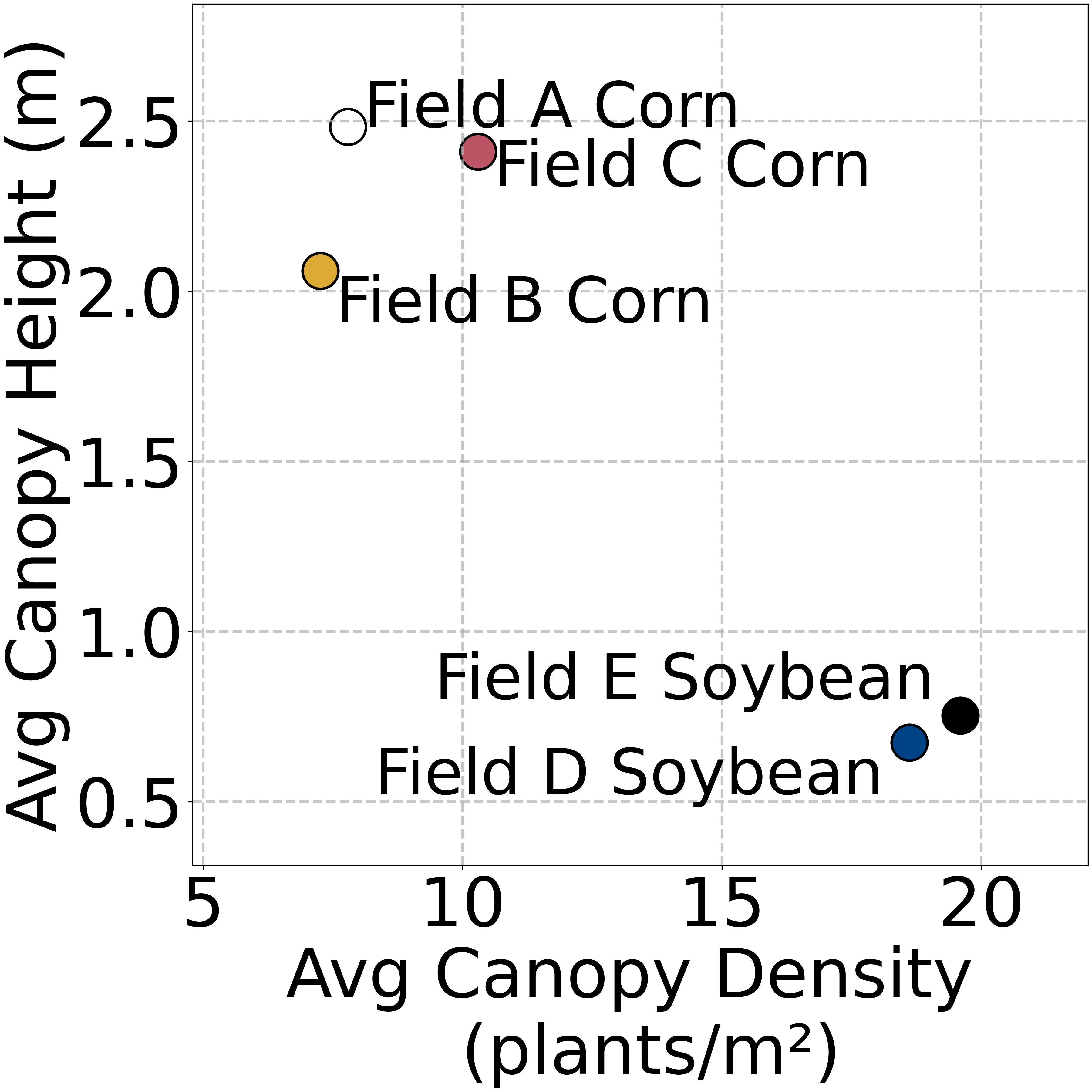}
\includegraphics[width=0.32\linewidth]{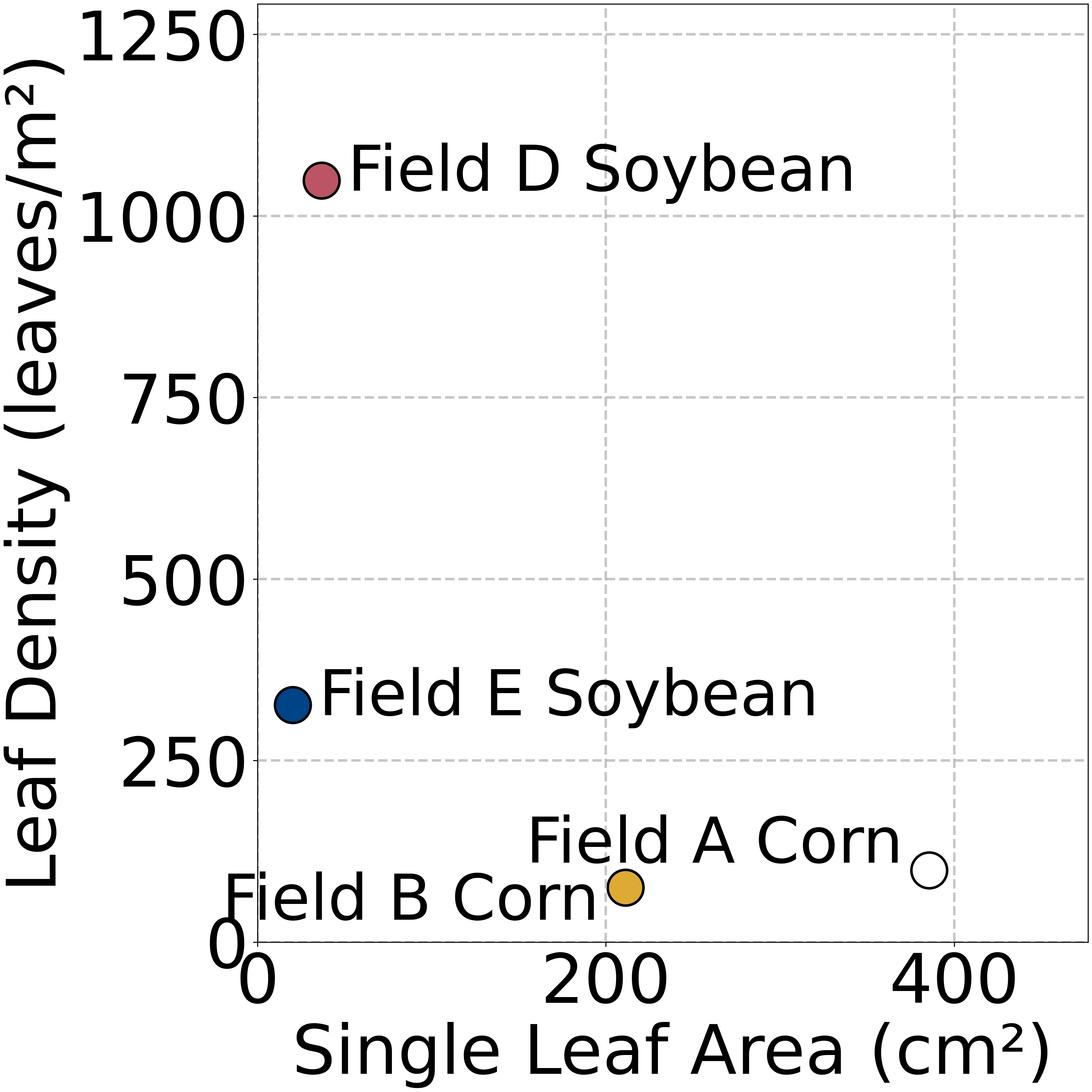}
\includegraphics[width=0.32\linewidth]{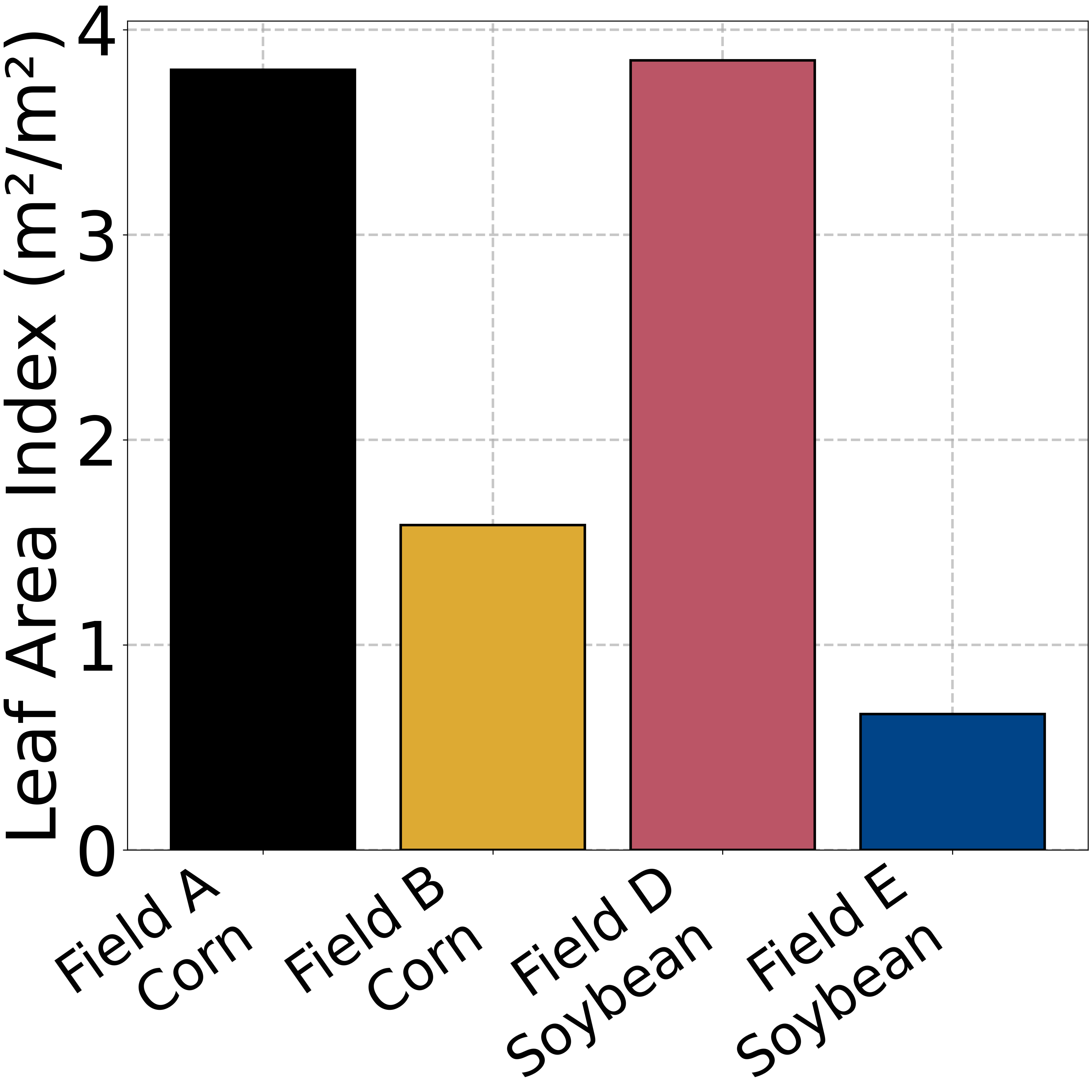}
 \caption{Distribution of canopy coverage parameters across fields.} 
\label{fig:groundtruth_scatter}
\end{figure}

We evaluated \sysname{} through an extensive in situ field campaign conducted across six agricultural fields spanning three canopy conditions: corn, soybean, and bare soil. Different planting times across fields led to varying canopy coverage and moisture levels, with soybean and corn fields exhibiting distinct degrees of canopy dryness. The distribution of groundtruth canopy parameters are shown in Figure~\ref{fig:groundtruth_scatter}. A total of 5 fields are covered with canopy, including 3 corn fields (A-C) and two soybean fields (D and E). The soil type across all fields is Drummer silty clay loam. Weather conditions varied throughout the 9-day campaign, including two rainfall events and several clear sunny days with maximum temperatures reaching 33°C, providing diverse environmental scenarios for evaluation. 

\change{\textbf{Radar and LiDAR Acquisition:} We collected radar data while hovering the UAV at multiple altitudes over multiple locations within each field. Each trial consisted of 100 radar scans (captured within a few seconds) at each UAV altitude (6 m and 8 m), recorded across 3–10 locations within each field. We collected LiDAR scans at both oblique and nadir incidence at an altitude of 12 m. In about 30 minutes, the UAV was able to scan a 4000 m$^2$ area at a point cloud density of nearly 5000 points/m$^2$. The GPR data acquisition and LiDAR scan can be performed simultaneously to save time.}


\textbf{Groundtruth Collection:} We collect groundtruth soil dielectric constant using the TEROS-12 capacitive sensor commonly used in similar research works \cite{josephson_low-cost_2021, khan_estimating_2022}. We record raw sensor values and use the manufacturer-provided conversion equation to convert to relative dielectric permittivity \cite{noauthor_teros_nodate}. At every TEROS-12 sampling location, we insert the probe into 3 random locations within 1 meter of the target location and average the resulting permittivity values. We verified the accuracy of the sensor by running a set of oven-dry experiments. 
We collect canopy structural groundtruth parameters using random sampling and tape measurements. Each of these measurements are averaged across 3 random sampling locations within each field. We also measure the gravimetric water content of the vegetation by drying it in an oven over 24 hours and measuring its mass before and after drying.



\begin{figure}[t]
    \centering
    \includegraphics[width=.9\linewidth]{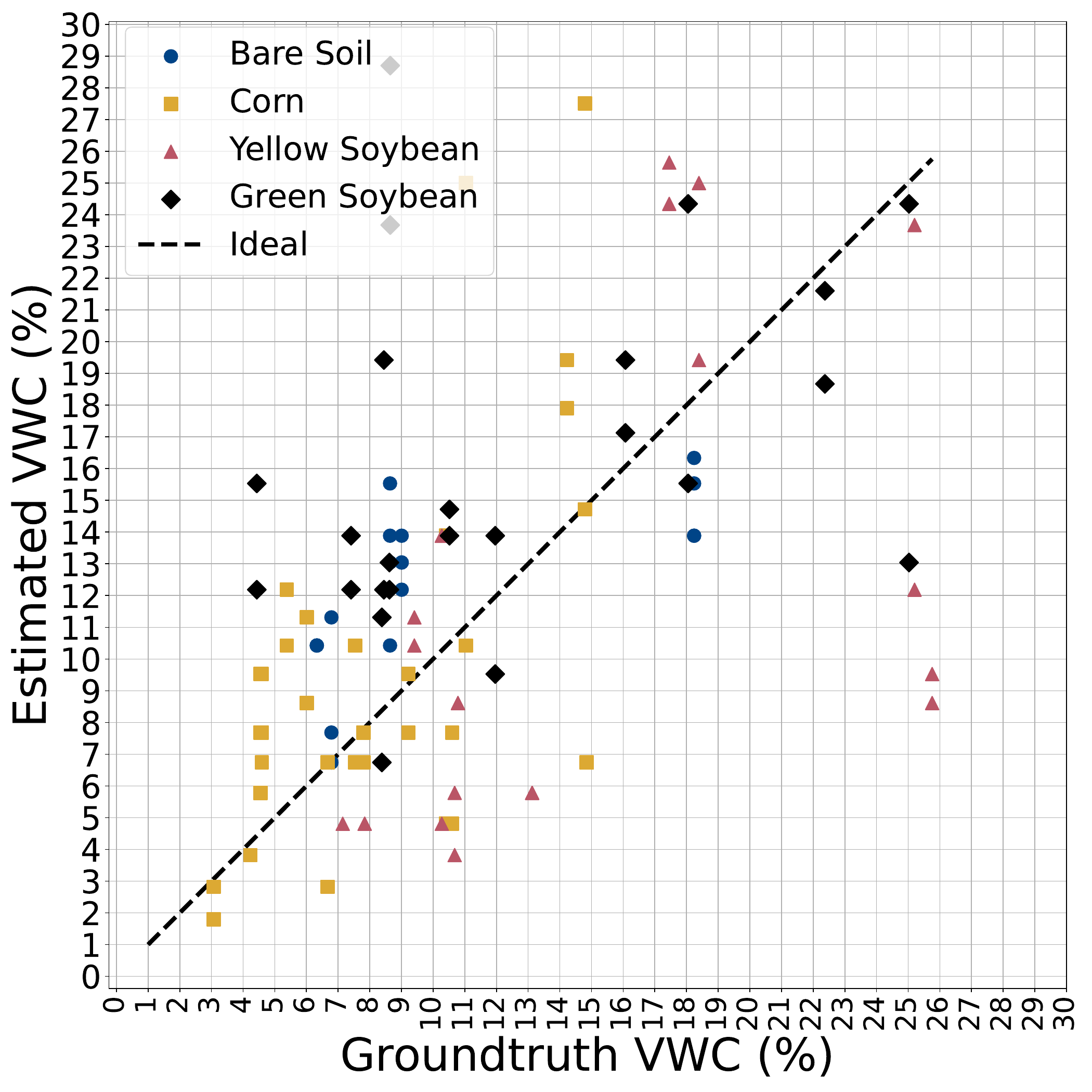}
    \caption{\sysname{} retrieval maintains positive estimation-groundtruth correlations over a wide range of groundtruth soil moistures.}
    \label{fig:eval_shocktest}
\end{figure}




\textbf{Evaluation Metrics:} We evaluate performance using the mean absolute error (MAE) between the estimated and groundtruth soil volumetric water content (VWC), which represents the ratio of the volume of water to the total volume of soil. We also use the canopy height, density, and leaf density parameters for evaluating the accuracy of our LiDAR-based canopy characterization.

\begin{figure}[t]
    \centering
    \includegraphics[width=.8\linewidth]{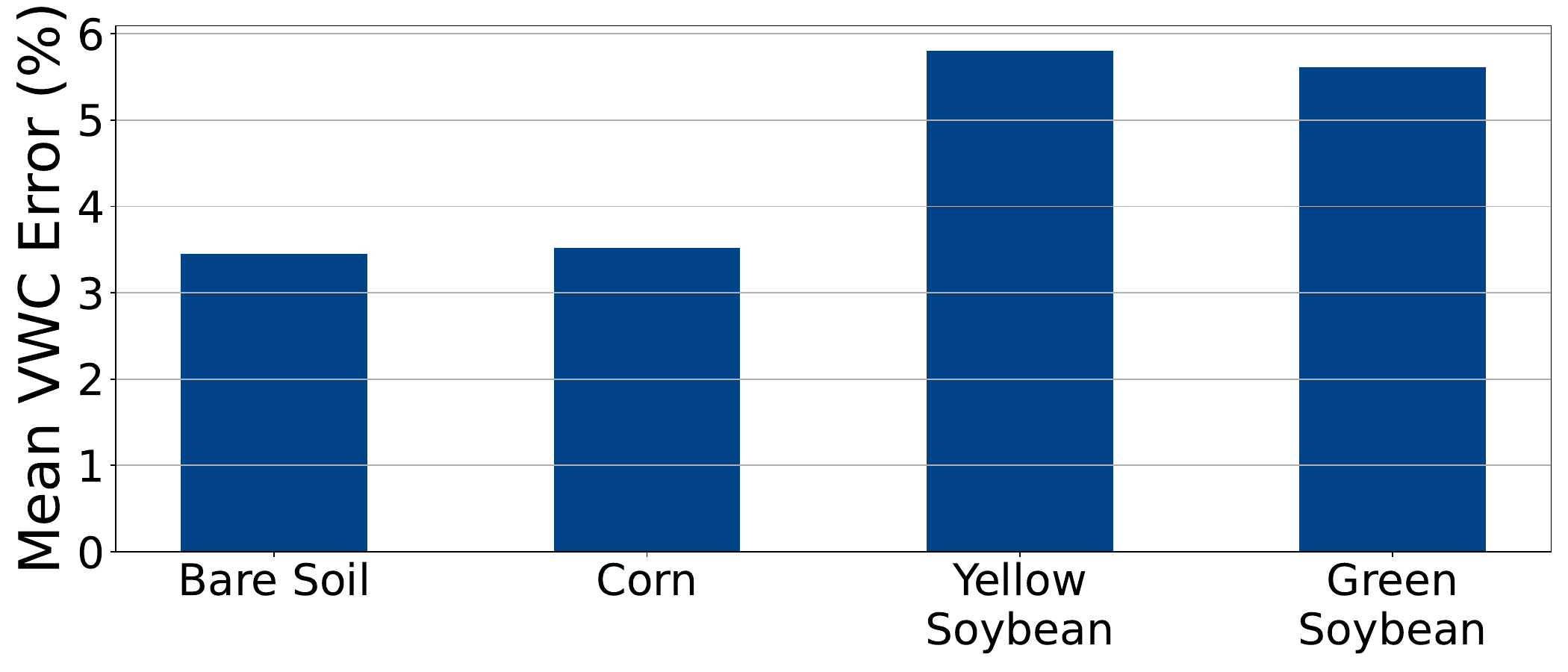}
    \caption{\sysname{} is robust in soil moisture sensing through various canopy covers.}
    \label{fig:eval_canopy_groups}
\end{figure}
\subsection{Overall Soil Moisture Sensing Accuracy}

All soil moisture retrievals were performed using data collected from a UAV \change{hovering in place over multiple locations within multiple agricultural fields}. \change{In all evaluation plots, where altitude or sampling location is not specified, moisture estimates were averaged over these dimensions}. \sysname{} maintains a positive correlation between estimated and ground-truth VWC across a wide range of soil moisture values (3–26\%) (Figure~\ref{fig:eval_shocktest}). Most of the error is concentrated in a small number of outlier points, primarily from yellow and green soybean fields. 
\change{An analysis of underestimated yellow and green soybean points ($\geq$25\% groundtruth VWC) shows unusually low RCS measurements across the whole GPR spectrum. This should be a result of the drone tilting a few degrees from nadir, resulting in partial ground reflection. Future work could filter out noisy measurements using the drone-mounted IMU data.}
\change{An analysis of overestimated green soybean and corn points (8-11\% groundtruth VWC) shows that these RCS measurements are unusually high at lower frequencies ($<$ 600 MHz), but closely align with the forward model at higher frequencies ($>$ 600 MHz). This effect may be explained by a subsurface reflection from a pocket of moisture, given the higher penetration of low frequency signals. This could be addressed in future work by modeling scattering from multiple soil layers.}

Figure~\ref{fig:eval_canopy_groups} demonstrates \sysname{} performance across different canopy environments. We can observe that \sysname{} achieves an average volumetric water content (VWC) error below 6\% regardless of the canopy coverage. Retrieval performance is higher over corn than soybean because the radiative transfer model more accurately represents corn with distinct stalk and leaf structures. In contrast, soybean canopies lack trunk-like stems, and the plants cannot be individually counted, resulting in lower accuracy when modeling soybean fields using only leaf-based canopy parameters.
\begin{figure}[t]
    \centering
    \includegraphics[width=\linewidth]{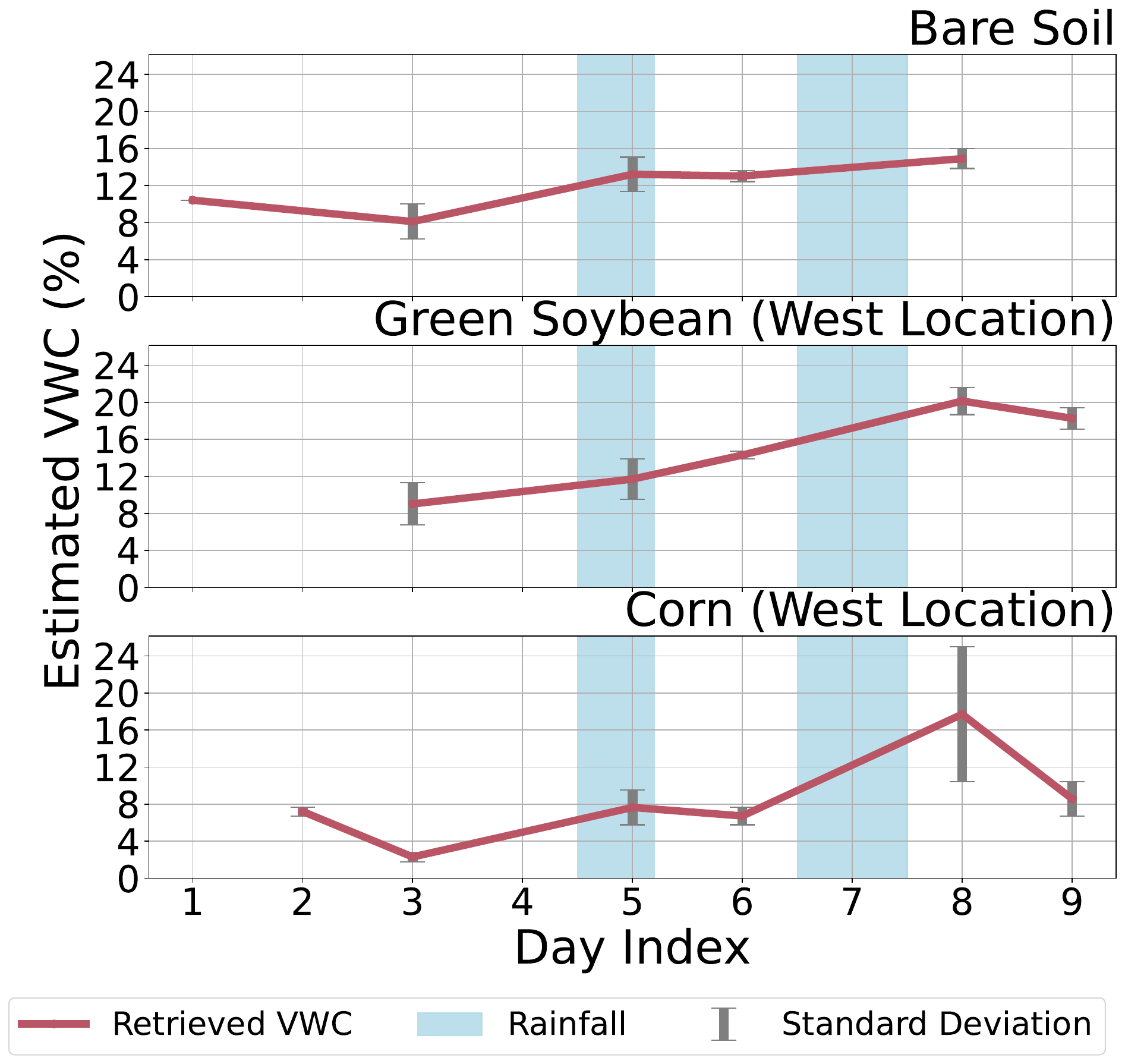}
    \caption{\sysname{} is capable of tracking daily soil moisture variations.}
    \label{fig:eval_longitudinal}
\end{figure}

\begin{figure*}[t]
    \centering
    \begin{minipage}{0.33\linewidth}
       \centering
    \includegraphics[width=\linewidth]{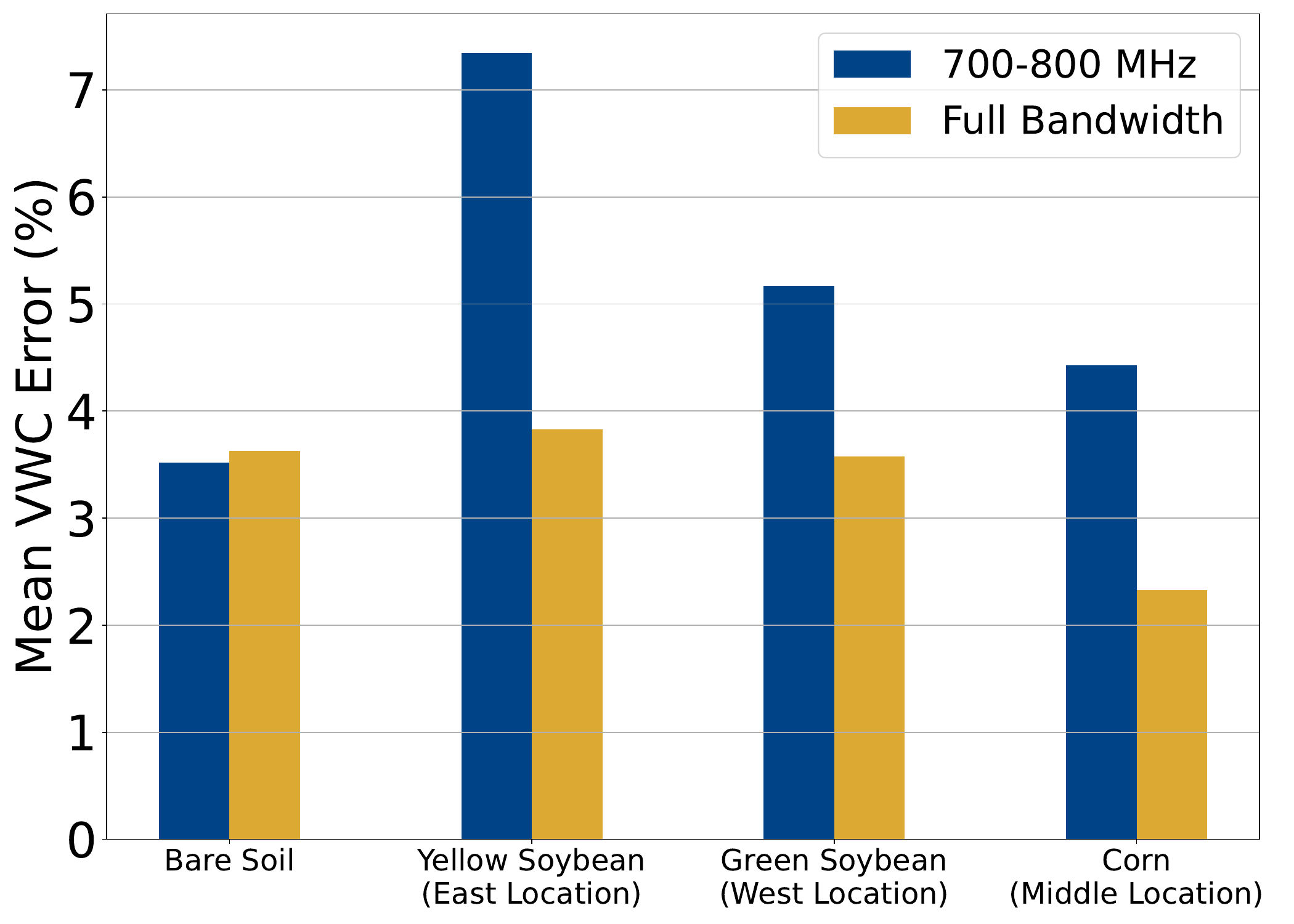}
    \captionsetup{width=.9\linewidth}

    \caption{\sysname{}'s soil moisture retrieval accuracy using full versus reduced radar bandwidth.}
    \label{fig:sensitivity_bandwidth}
    \end{minipage}\hfill
    \begin{minipage}{0.33\linewidth}
         \centering
         \captionsetup{width=.9\linewidth}
    \includegraphics[width=\linewidth]{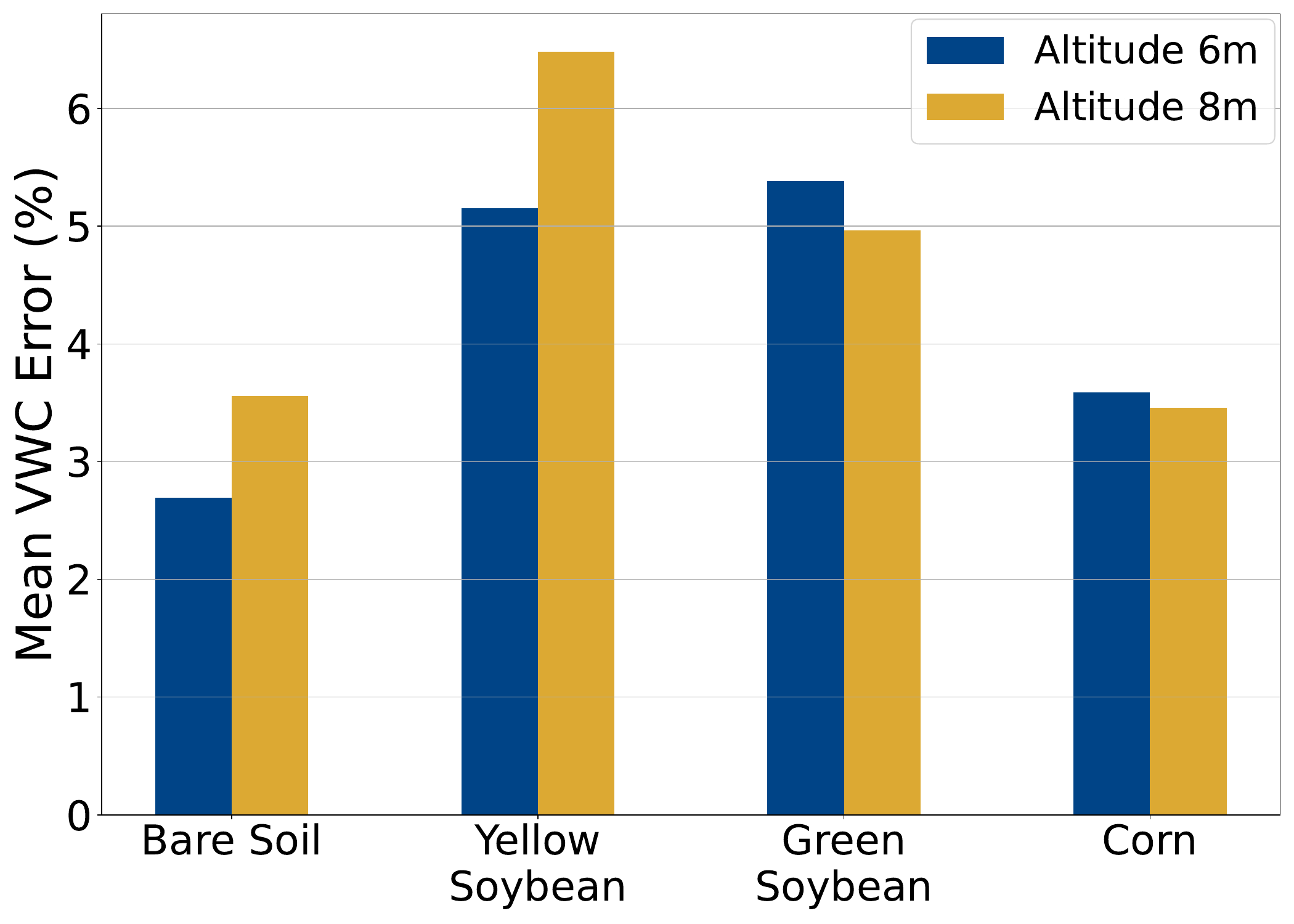}
    \caption{\sysname{} remains robust to UAV altitudes by accounting for the effective radar footprint.}
    \label{fig:sensitivity_altitude}
    \end{minipage}
    \begin{minipage}{0.33\linewidth}
        \centering
        \captionsetup{width=.9\linewidth}
    \includegraphics[width=\linewidth]{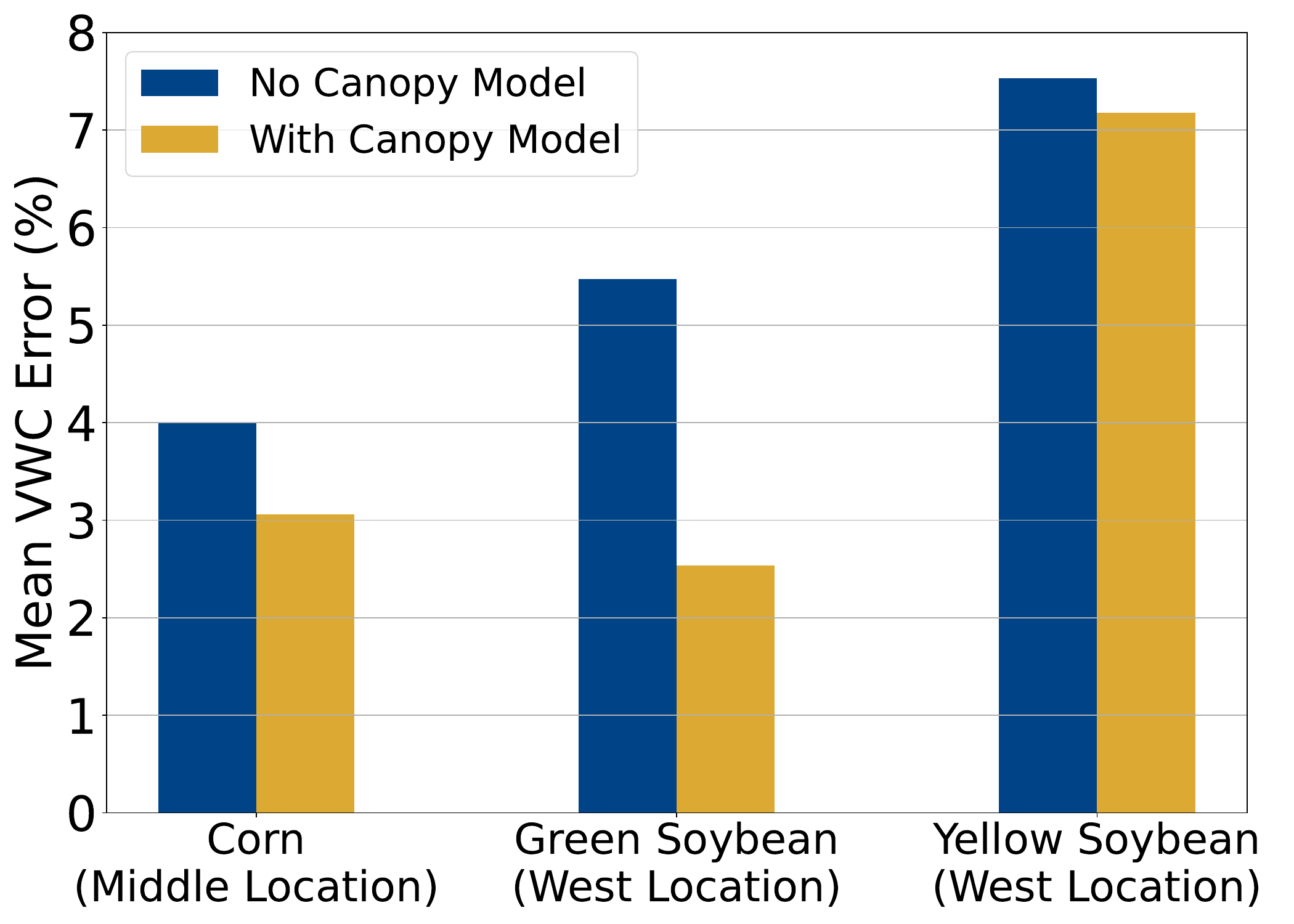}
    \caption{Effectiveness of \sysname{} radiative transfer model in decoupling canopy and soil contributions.}
    \label{fig:sensitivity_canopy}
    \end{minipage}
\end{figure*}

\vspace{.3em}\noindent \textbf{Longitudinal Performance:}
As shown in Figure~\ref{fig:eval_longitudinal}, \sysname{} successfully tracks temporal variations in soil moisture, with estimated VWC increasing during rainfall events across all canopy types. This capability demonstrates that the system can capture soil moisture dynamics over time, supporting daily irrigation planning and improved field-scale water management. \change{The large standard deviation on the corn soil moisture estimate of day 8 is due to a large spread in the RCS measurements collected at this time and location. We attribute this spread of measurements to variations in the orientation of the radar, which can occur during windy weather. Future work can account for radar orientation in the radiative transfer model using on-board UAV sensors. }


\subsection{Sensitivity Analysis}


\change{We perform a series of sensitivity analyses to better understand the impact of different variables.}

\vspace{.3em} \noindent \textbf{Impact of Radar Bandwidth: } \sysname{} leverages the radar’s wide operational bandwidth to capture the nonlinear frequency response of both soil and vegetation, which is essential for accurate soil moisture retrieval. The radar cross section (RCS) of the ground and the attenuation from the canopy are both frequency dependent, reflecting variations in dielectric and scattering properties across the band. In contrast, drone-related factors such as altitude or orientation shifts manifest as nearly constant offsets across frequencies and can be separated from the frequency-dependent components of the signal. To evaluate the effect of radar bandwidth, we compared \sysname{}’s retrieval accuracy using only the top 100 MHz of the frequency response with that obtained from the full bandwidth. As shown in Figure~\ref{fig:sensitivity_bandwidth}, retrieval with the reduced bandwidth performs worse in the selected locations. These results demonstrate that exploiting the full frequency range provides richer canopy and soil information distributed across the RCS spectrum, leading to more reliable and accurate soil moisture estimation under diverse canopy conditions. \change{We can see that the bare soil VWC error has minimal improvement using a wider bandwidth, which emphesizes the importance of bandwidth for canopy and soil decoupling.}

\vspace{.3em} \noindent \textbf{Impact of UAV Altitude: } We collected radar measurements at two UAV altitudes, 6 m and 8 m, to evaluate altitude-dependent effects on retrieval accuracy. As shown in Figure \ref{fig:sensitivity_altitude}, \sysname{}'s retrieval performance remains consistent within 1.5\% across both altitudes, indicating that it can accurately compensate for variations in flight height. An increase in altitude naturally reduces the received ground reflection power due to free-space path loss, which can degrade the performance of reflection amplitude–based soil sensing methods that do not explicitly correct for this effect.
\sysname{} incorporates altitude corrections directly in both the RCS measurement calculation (Equation~\ref{eq:measured_rcs}) and the RCS simulation (Equation~\ref{eq:effective_reflective_area}), ensuring that the retrieved RCS values remain physically consistent across altitudes. The results in Figure~\ref{fig:sensitivity_altitude} demonstrate that the measured RCS scales proportionally with the radar footprint on the ground. Furthermore, the agreement between measurements at different altitudes indicates that the small-footprint approximation used in our model is valid, allowing \sysname{} to maintain reliable soil moisture retrieval even when UAV flight height varies during data collection.



    

   

    

\begin{figure}
    \centering
    \includegraphics[width=\linewidth]{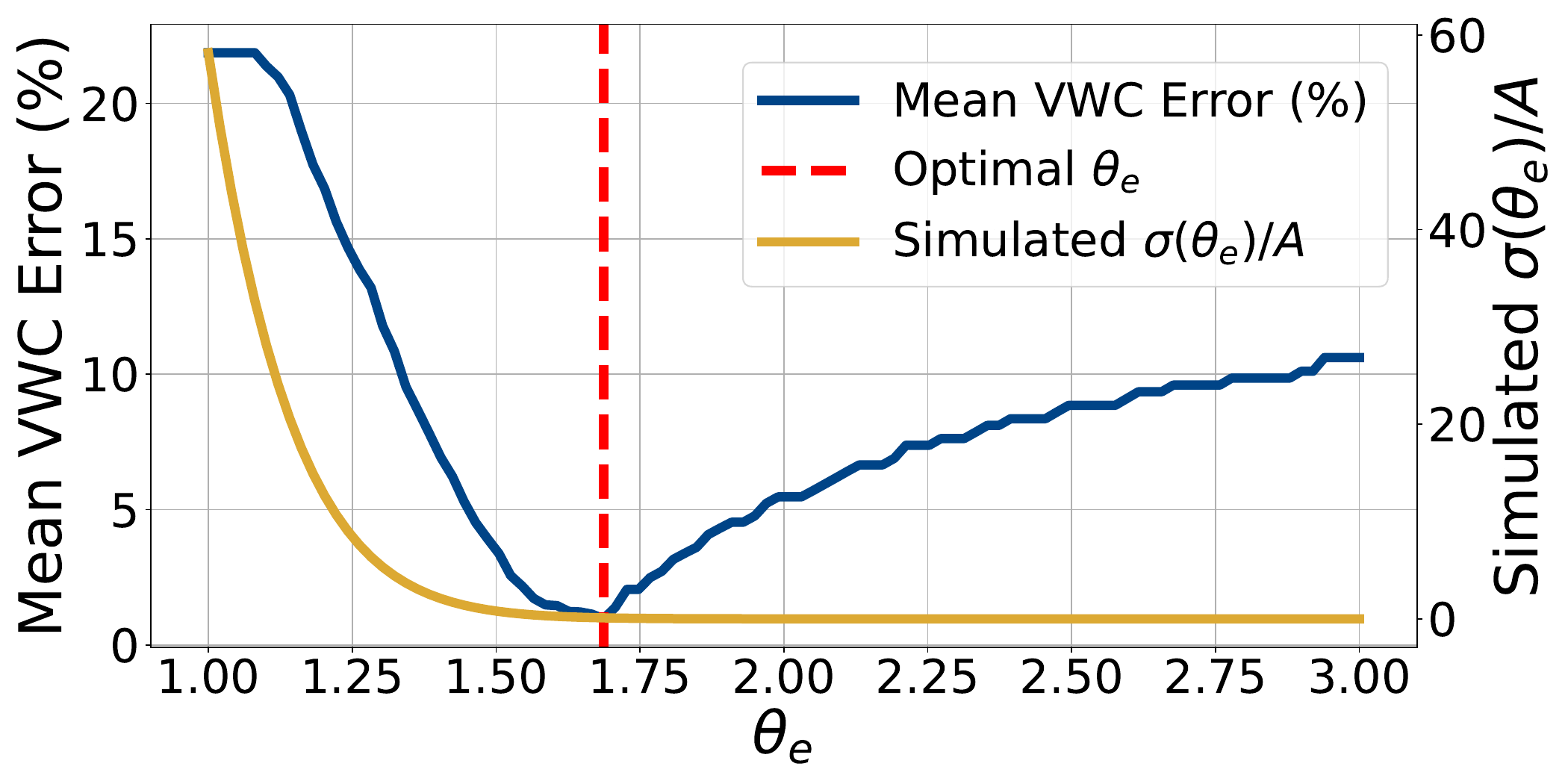}
    \caption{\sysname{} achieves the lowest soil moisture retrieval error when the effective radar beamwidth aligns with the angular range of the coherent ground reflection, validating the small-footprint approximation}
    \label{fig:justify_small_footprint_bw_tuning}
\end{figure}

\vspace{.3em} \noindent \textbf{Ablation Study - Canopy Modeling: }
To evaluate the effectiveness of canopy modeling in improving soil moisture retrieval, we conducted an ablation study using two fields with dense vegetation canopies. \sysname{} jointly estimates soil and canopy permittivity within the radiative transfer model, allowing it to decouple canopy attenuation effects from soil backscatter. To demonstrate the impact of this capability, we compared retrieval results with and without canopy modeling enabled.
As shown in Figure~\ref{fig:sensitivity_canopy}, including canopy modeling improves retrieval accuracy \change{most significantly in the Green Soybean field, which shows a two-fold improvement in accuracy at the selected location. This is a reasonable result, as the gravimetric water content of the green soybean plants was higher than the yellow soybean and corn plants, suggesting a higher attenuation was applied to the ground reflection in this location.} This result highlights not only the importance of explicitly modeling canopy effects in radar-based soil moisture estimation but also the robustness of \sysname{} in effectively separating and characterizing the coupled soil–canopy interactions present in real-world agricultural environments.

\vspace{.3em} \noindent \textbf{Impact of Effective Radar Beamwidth: } In \sysname{}, we adopt a small-footprint approximation within the radiative transfer model to account for the dominant coherent component of the ground reflection. Instead of modeling radar interactions across the entire antenna 3 dB beamwidth, we compute an effective radar beamwidth that captures only the portion of the footprint contributing coherently to the received signal. This approach simplifies computation while preserving physical accuracy, as the coherent component dominates the backscattered energy in nadir-looking radar configurations.
To evaluate the validity of this assumption, we manually varied the effective beamwidth parameter and examined its impact on soil volumetric water content (VWC) retrieval accuracy. As shown in Figure~\ref{fig:justify_small_footprint_bw_tuning}, the minimum retrieval error occurs at the effective beamwidth $\theta_\text{e}$ where the ground RCS begins to decay with increasing incidence angle. This correspondence indicates that our empirically tuned effective beamwidth aligns with the angular extent of the coherent component, confirming that the small-footprint approximation in \sysname{} accurately captures the dominant scattering behavior.

\begin{figure}[t]
    \centering
    
    \begin{subfigure}{\linewidth}
        \centering
        \includegraphics[width=.99\linewidth]{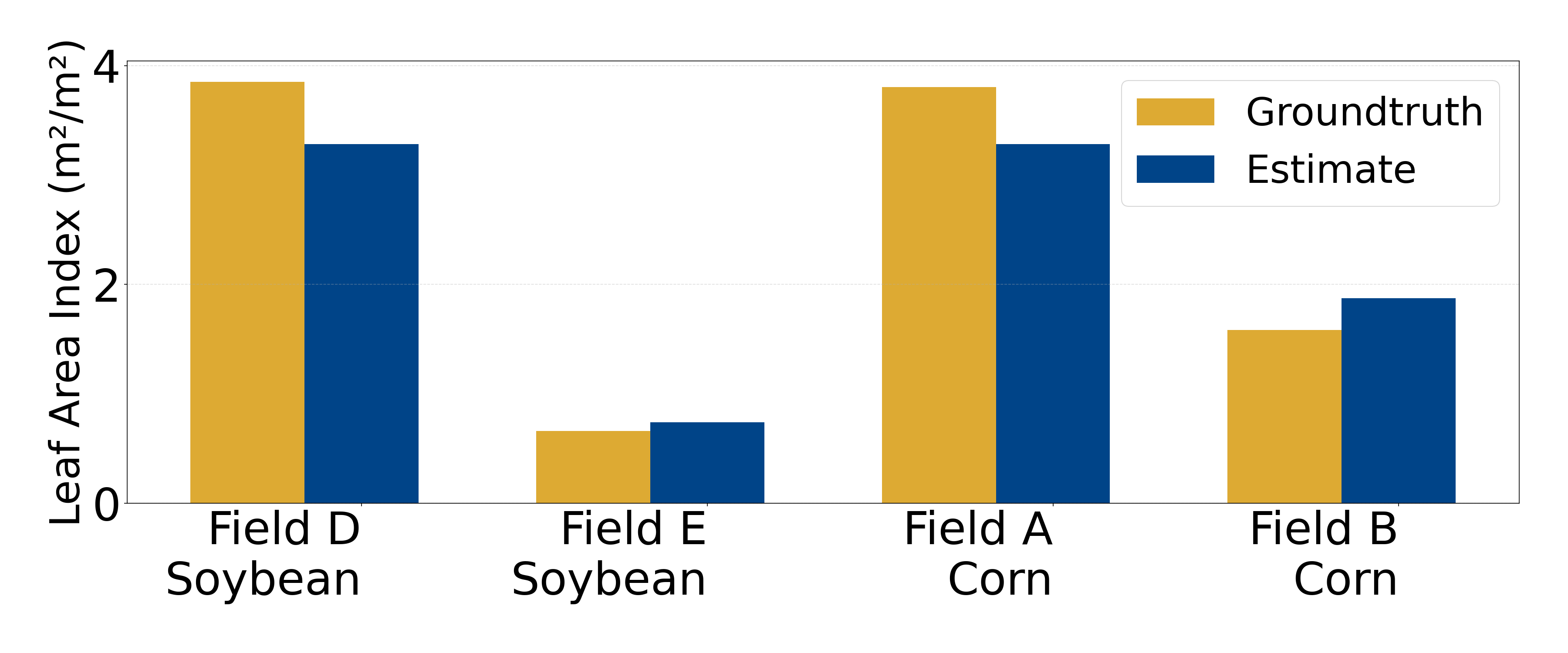}
        \caption{} 
    \end{subfigure}

    \begin{subfigure}{\linewidth}
        \centering
        \includegraphics[width=.99\linewidth]{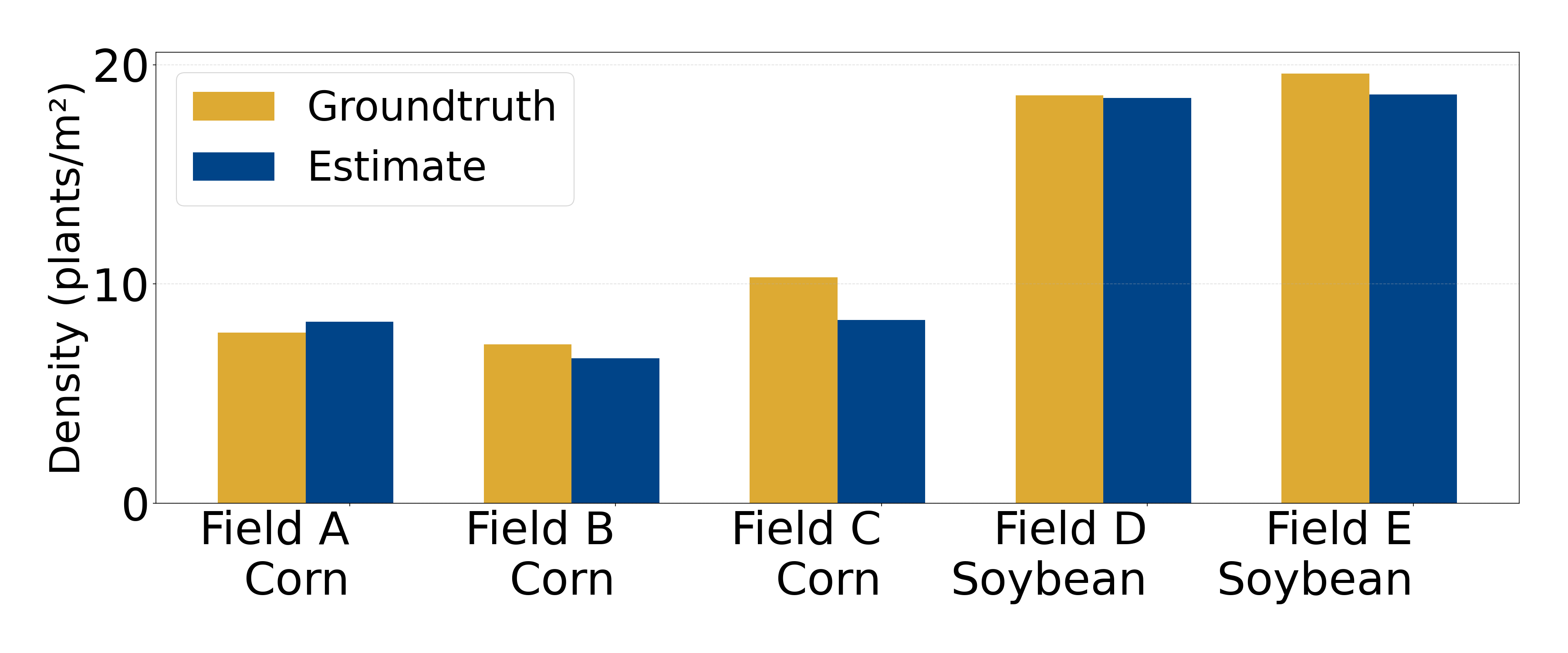}
        \caption{} 
    \end{subfigure}

    \begin{subfigure}{\linewidth}
        \centering
        \includegraphics[width=.99\linewidth]{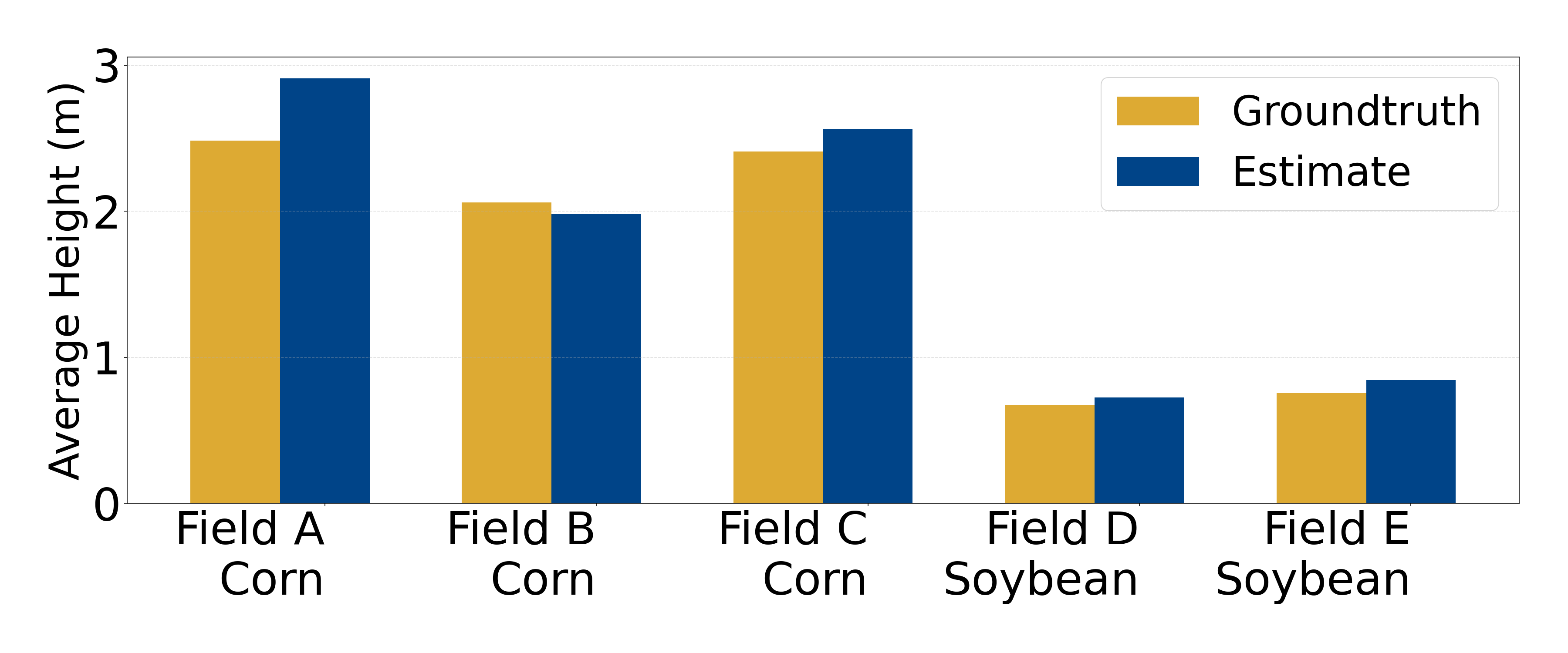}
        \caption{} 
    \end{subfigure}

    \caption{Comparison of LiDAR-derived estimates and ground-truth: 
    (a) Leaf Area Index,  
    (b) Average Canopy Density, and
    (c) Average Canopy Height.}
    \label{fig:eval_all_canopy_parameter_barplots}
\end{figure}
\subsection{LiDAR-Based Canopy Parameter Estimation Performance}
We evaluated the accuracy and generalization capability of our canopy parameter estimation pipeline using UAV LiDAR scans collected over five agricultural fields with ground-truth canopy measurements. Figure~\ref{fig:eval_all_canopy_parameter_barplots} compares the LiDAR-derived canopy parameters with their corresponding field-measured values.

Our results show that the LiDAR-based model preserves high accuracy across fields with varying crop type, canopy height, and planting density. Height estimation remains consistent with an average relative error of 8.92\%, and estimated plant density follows the ground-truth trend with an average relative error of 8.19\%. These findings indicate that \sysname{}’s canopy model generalizes effectively across diverse canopy architectures without requiring field-specific tuning.

We also evaluated leaf area index (LAI) estimation across test fields using our voxel-based inversion method. As shown in Figure~\ref{fig:eval_all_canopy_parameter_barplots}, the LiDAR-derived LAI closely tracks the field-measured values, achieving an average error of 14.71\%. The consistent performance across different canopy conditions demonstrates that \sysname{} robustly captures key vegetation structural parameters essential for accurate radiative transfer modeling.




%% file: sections/6-discussion.tex

\section{Discussion}
\change{In the previous section, we demonstrated that \sysname{} delivers through-canopy soil moisture sensing across a variety of challenging conditions. These properties make \sysname{} a natural fit for large-scale agricultural applications. Several aspects of \sysname{}'s design can be further improved to better support all real-world conditions at scale.}




\vspace{.3em} \noindent  \change{\textbf{Subsurface Soil Moisture Effects: } 
In this study, we focus on surface-level scattering and do not explicitly model subsurface contributions. However, prior work~\cite{khan_estimating_2022} shows that soil moisture gradients can generate radar returns from subsurface layers, particularly at lower frequencies (200–500 MHz) where penetration depth increases. In our future work, we will extend the radiative transfer framework to incorporate multilayer dielectric stratification and frequency-dependent attenuation, enabling inversion of depth-dependent moisture profiles from surface backscatter.}


\vspace{.3em} \noindent \change{\textbf{Estimation of Soil Moisture Depth Profiles: } While \sysname{} senses only surface soil moisture, it may be possible to estimate soil moisture depth profiles by using a time-series of surface measurements to initialize a physics-based hydrological model \cite{walker_one-dimensional_2001}. More specifically, surface soil moisture estimates from \sysname{} could initialize the top boundary condition by which the Richards fluid flow equation could be numerically solved.}



\vspace{.3em} \noindent \change{\textbf{Drone Mobility and Soil Moisture Mapping: } Our current implementation of \sysname{} evaluates soil moisture retrieval accuracy at individual radar footprints due to limitations in available ground-truth data. While these measurements demonstrate reliable performance within each footprint, future work will extend the system to operate during drone motion, enabling continuous soil moisture estimation and large-scale map generation across the surveyed area.}

\vspace{.3em} 
\noindent 
\change{ \textbf{Soil Roughness Modeling: } 
In \sysname{}, soil surface roughness is estimated empirically at the beginning of the planting season and treated as a fixed parameter in the radiative transfer model. Future work could improve retrieval accuracy by incorporating LiDAR-derived elevation data prior to canopy emergence to directly estimate roughness. The framework could also be extended to jointly infer soil roughness together with soil and canopy permittivity, reducing reliance on external calibration.}

\vspace{.3em} \noindent
\change{\textbf{Generalization to Diverse and Complex Canopies: } 
While \sysname{} was evaluated on corn and soybean canopies, its radiative transfer framework can be extended to other vegetation types, including woody plants and structurally complex canopies such as forests. Woody vegetation can be modeled using dielectric cylinder representations similar to row crops, whereas succulent or thick-leaf plants may require decomposition into discrete scatterers with appropriate scattering matrices. Highly heterogeneous canopies introduce additional challenges due to irregular geometry, and spatial variability, motivating future development of spatially adaptive radiative transfer models that capture local canopy structure and align with LiDAR footprints across the sensing area.}

%% file: sections/7-conclusion.tex
\section{Conclusion}
\sysname{} enables reliable UAV-based soil moisture sensing through fully-grown crop canopies. Our methodologies combine nadir wideband radar modeling, a canopy-aware radiative transfer model, and canopy parameter estimation pipeline from UAV LiDAR scans. By modeling coherent ground reflections and canopy attenuation using physically-grounded scattering principles, our system retrieves soil permittivities with the wide-band ground reflection captured by the radar. Field experiments over corn, soybean, and bare soil demonstrate consistent performance, and \sysname{} achieves low volumetric water content error on average (\change{4.49\%}) across diverse canopy conditions. The results show that our system provides a robust path for practical, physics-based, through-canopy soil moisture sensing.